\documentclass{article}
\usepackage{amssymb,amsmath}
\newcommand{\p}{{\partial}}
\newcommand{\PSbox}[3]{\mbox{\rule{0in}{#3}\includegraphics{#1}\hspace{#2}}}

\newtheorem{thm}{Theorem}
\newtheorem{cor}[thm]{Corollary}
\newtheorem{lemma}[thm]{Lemma}
\newtheorem{prop}[thm]{Proposition}

\newenvironment{proof}{{\bf Proof. }}{\hfill$\square$ \vskip .5cm}

\newcommand{\R}{{\mathbb R}}
\newcommand{\C}{{\mathbb C}}
\newcommand{\Z}{{\mathbb Z}}

\newcommand{\E}{{\mathbb E}}
\newcommand{\Bo}{{\mathbf B}_0\!}
\newcommand{\Bl}{{\mathbf B}_1\!}
\newcommand{\dbar}{\partial_{\overline z}}
\renewcommand{\Re}{\mbox{Re}}
\renewcommand{\Im}{\mbox{Im}}
\newcommand{\eps}{{\epsilon}}
\newcommand{\vep}{{\varepsilon}}

\begin{document}
\title{Conformal invariance of domino tiling}
\author{Richard Kenyon
\thanks{CNRS UMR8628, Laboratoire de Topologie, B\^at. 425,
Universit\'e Paris-Sud, 91405 Orsay, France. 
This work was begun while the author was at
CNRS UMR128, \'Ecole Normale Sup\'erieure de Lyon, Lyon, France.}}
\maketitle

\abstract{
Let $U$ be a multiply-connected region in ${\bf R}^2$
with smooth boundary. Let $P_\epsilon$ be a polyomino
in $\epsilon{\bf Z}^2$ approximating $U$ as $\epsilon\to0$. We show that,
for certain boundary conditions on $P_\epsilon$,the height distribution on a random domino tiling (dimer
covering) of $P_\epsilon$ is conformally invariant
in the limit as $\epsilon$ tends to $0$,
in the sense that the distribution of heights
of boundary components (or rather,
the difference of the heights from their mean values)
only depends on the conformal type of $U$.
The mean height is not strictly conformally invariant
but transforms analytically under conformal mappings in a
simple way. The mean height
and all the moments are explicitly evaluated.
}
\medskip

\noindent key words: Domino tilings, conformal invariance

\section{Introduction}
Conformal invariance of a lattice-based statistical mechanical system 
is a symmetry property of the system at large scales.
It says that, in the limit as the lattice spacing $\eps$ tends to $0$,
macroscopic quantities associated with the system transform 
covariantly under conformal maps of the domain. 

\begin{figure}[htbp]
\begin{center}
\PSbox{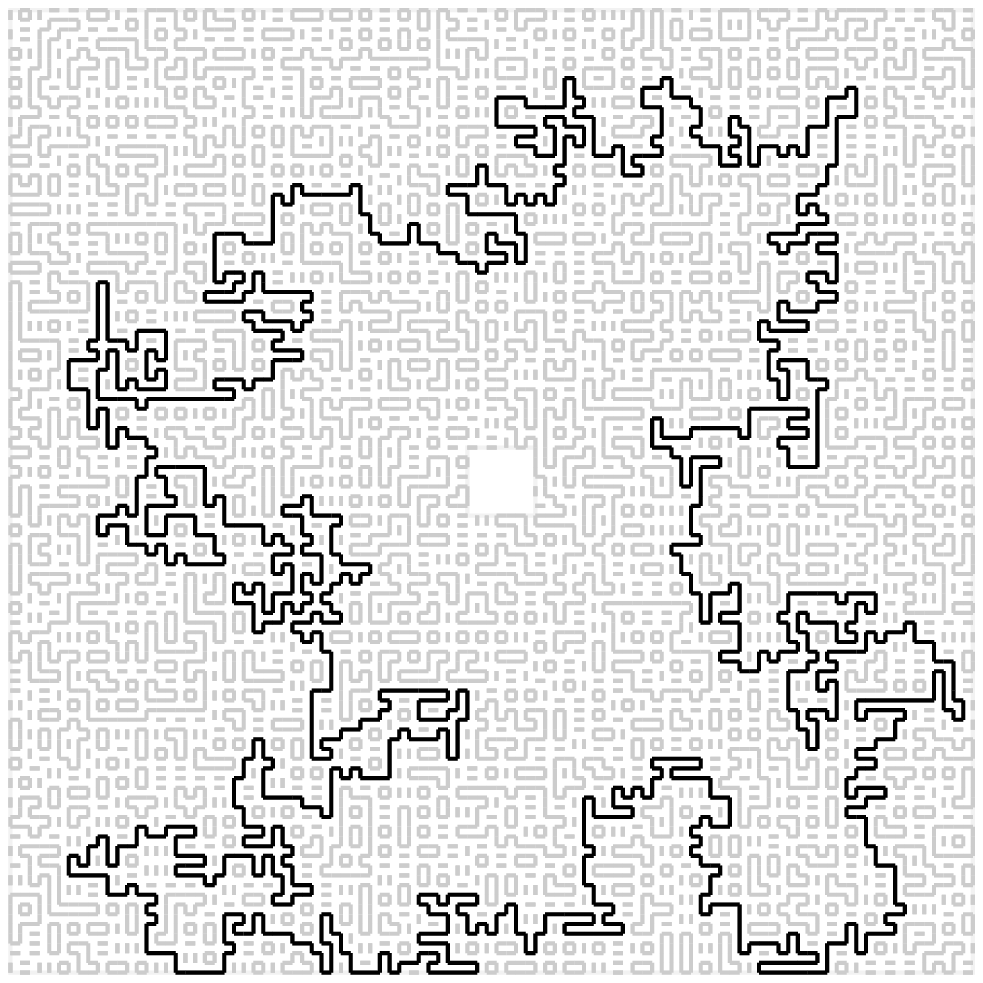}{1in}{3.7in}
\end{center}
\caption{\label{annulus}A cycle in a union of two random
domino tilings of an annulus.}
\end{figure}

Conformal invariance for statistical mechanical lattice
models is a physical principle which until now has not been
proved except in certain models which were tailored to be conformally
invariant \cite{Cardy} (recently in \cite{BS}
Benjamini and Schramm
prove conformal invariance in a discrete, but non-lattice, percolation model). 
Nonetheless conformal invariance is an extremely powerful principle:
in the plane, conformally invariant models are classified, in a sense,
by representations of the Virasoro algebra \cite{BPZ}.
Physicists have
used this theory fruitfully to compute exact ``critical exponents" and other physical
quantities associated to critical lattice models \cite{Cardy}.
For example, the cycle in Figure \ref{annulus} is believed to have Hausdorff
dimension $\frac32$ in the limit (see e.g. \cite{KH}) 
and the path in Figure \ref{treefig} is believed to have dimension $\frac54$
\cite{Gutt}. 
Although many well-known models are believed to be conformally invariant
at their critical point, no rigorous techniques
were known to prove conformal invariance in these models.

In this paper we deal with the two-dimensional lattice dimer model, or domino
tiling model
(a domino tiling is a tiling with $2\times 1$ and $1\times 2$ rectangles).
We prove that in the limit as the lattice
spacing $\eps$ tends to zero, certain macroscopic 
properties of the tiling are conformally invariant.

The {\bf height function} $h$ on a 
domino tiling is an integer-valued function on the 
vertices in a tiling. It is defined below in section \ref{ht};
see also \cite{BH,Thurston}.
One can think of a domino tiling of $U$ as a map $h$ from $U$
to $\Z$, where for each unit lattice square, the images of the four vertices
under $h$ are $4$ consecutive integers $v,v+1,v+2,v+3$. Furthermore
each boundary edge of $U$ must have image of length $1$ and not $3$.
The map $h$ defines and is defined by the tiling: the edges crossed by a domino
are those whose image under $h$ has length $3$.  Our main result is
the conformal invariance of $h$ for a random tiling:

\begin{thm}\label{1}
Let $U$ be a bounded, multiply connected domain in $\C=\R^2$ with 
$k+1$ smooth boundary components, 
each with a marked point $d_0,d_1,\ldots,d_k$.
Let $\{P_{\epsilon}\}_{\eps>0}$ be a sequence of
polyominos, with $P_\eps\in\epsilon\Z^2$, approximating $U$
as described in section \ref{approx}.
Let $d_j^{(\eps)}$ be a vertex of $P_\eps$ within $O(\eps)$ of $d_j$.
Let $\mu_\epsilon$ be the uniform measure on domino tilings
of $P_\epsilon$. Then the joint distribution of the height variations
of the points $d^{(\eps)}_j$ (that is, the difference of the heights
from their mean value) tends to a finite limit which is conformally invariant.
\end{thm}

By conformal invariance we mean, if $f\colon U\to U'$ is a conformal isomorphism
then the distribution of the height variations of $f(d_j)$ is the same
as the distribution of the height variations of the $d_j$ themselves.

The mean height of a point of $P_\eps$ is not strictly
conformally invariant in the limit: there is an extra term coming from the
heights on the boundary (Theorem \ref{meanht}). We prove there that
the limiting mean height is a harmonic function on $U$ whose boundary
values depend on the tangent direction of the boundary.

The picture of the height function is completed by understanding the 
distribution of heights at interior points of $U$. For an interior
point $x$ of $P_\eps$, Theorem \ref{0} below
and \cite{Kenyon} show that the height $h(x)$ 
tends to a Gaussian with variance
$c\log(\frac1\eps)$ for a constant $c$ (which can be shown to be $\frac8{\pi^2}$
by a computation similar to that in \cite{Kenyon}). See below.
This variance diverges as $\eps\to0$. On the other hand
the proof of Theorem \ref{1} shows that the moments 
$$\E((h(x_1)-\overline{h(x_1)})(h(x_2)-\overline{h(x_2)})\cdots (h(x_m)-
\overline{h(x_m)}))$$
for distinct $x_i$ tend to a finite and conformally invariant limit.

Theorem \ref{1} can be extended to regions $U$ with piecewise smooth boundary,
on condition that at each corner the boundary tangents have one-sided limits. See below.
\medskip

Figure \ref{annulus} illustrates one consequence of Theorem \ref{1}.
In that figure we took two random 
domino tilings of an annular region (a square with a square hole).
A domino tiling corresponds to
a dimer covering, or perfect matching, of the underlying graph 
(a perfect matching is a collection of edges covering each vertex exactly once).
Two perfect matchings form a union of closed cycles and doubled edges
in the graph. 
One can ask about the distribution of the number of cycles separating
the inner and outer boundaries of the annulus (there is just one 
such cycle in the figure).
The argument of \cite{Kenyon} shows that the distribution of the height
difference between two boundary components for a single domino tiling
is directly related to the 
distribution of the number of cycles separating those two components
in a union of two tilings.
Indeed, the expected number of cycles is $\frac1{16}$ times
the variance of the height difference.
Theorem \ref{1} therefore implies that
the distribution of the number of cycles separating the boundary components
from each other is conformally invariant.

Another interpretation of the height function uses the connection
between domino tilings and spanning trees on $\Z^2$ \cite{BP}. 
In section \ref{trees} we relate the height function to 
the ``winding number" of arcs in the corresponding spanning tree.
\bigskip

Theorem \ref{1} follows from a more fundamental result.
The {\bf coupling function} on $P_\eps$ 
is a function $C\colon P_\eps\times P_\eps\to\C$
which determines the measure $\mu_\eps$ 
(the uniform measure on the set of all tilings of $P_\eps$) in the sense that
subdeterminants of the coupling function matrix give probabilities
of finite configurations of dominos occurring in a tiling \cite{Kenyon}.
The coupling function is closely related to the Green's
function.
The following is a loose statement of the result.

\begin{thm}\label{0} Let $U$ and $\{P_\eps\}_{\eps>0}$
be defined as in Theorem \ref{1}. Let $v\neq w$ be points
in the interior of $U$ and
$v^{(\eps)},w^{(\eps)}$ vertices of $P_\eps$ within $O(\eps)$ of $v,w$
respectively.
The coupling function $C$ for domino tilings of $P_\eps$ satisfies
$$C(v^{(\eps)},w^{(\eps)})= \eps F_j(v,w) + o(\eps),$$
where $j=0$ or $1$ depending on a parity condition, where
$F_0$ and $F_1$ are analytic in the second variable and
depend only on the conformal type of $U$.
\end{thm}

For a precise statement see Theorem \ref{longrange}.
This result has an immediate corollary regarding densities of local configurations.
\begin{cor}\label{deviation} In a random tiling of $P_\eps$,
the expected density of occurrences of a local configuration $E$ of dominos
at a point $v$ in the interior of $U$
is of the form $c(E) + \eps W_E(v)+o(\eps)$, where $c(E)$ equals
the density of $E$ in a random tiling of the whole plane $\eps\Z^2$, and 
$W_E$ is a function depending only on the conformal type of $U$.
\end{cor}

The proofs of the above results are given for polyominos with somewhat special boundary conditions.
We discuss in section \ref{conclusion} alternate boundary conditions for which
it may be possible, using similar methods, to prove similar results.
We remark that certain restrictions on the boundary are definitely necessary, however:
in \cite{CKP} Cohn, Kenyon and Propp compute the mean height
when the height function on the boundary is of order $\frac1\eps$.
In this case the mean height satisfies a much more complicated non-linear
elliptic PDE and does not appear to have any simple conformal invariance properties.
\medskip

The paper is organized as follows. 
In section \ref{graphdefs} we define the polyominos, 
graphs and notations we will be using.
We also define the height function.
In section \ref{DAF} we
define discrete analytic functions, and show that the coupling function is one.
In section \ref{boundary} we discuss boundary values of
the coupling function.
In section \ref{limits} we prove Theorem \ref{0}.
In section \ref{1proof} we prove Theorem \ref{1} using Theorem \ref{0}.
In section \ref{avght} we compute explicitly the average
height function on a region.
In section \ref{trees} we discuss the connection with spanning trees,
and in section \ref{conclusion} we discuss other boundary conditions
and give some concluding remarks.
\medskip

\noindent{\bf Acknowledgements.}
I would like to thank Oded Schramm for many helpful ideas, and the referee for several simplifications
in section \ref{1proof}.

\section{Definitions}\label{graphdefs}
\subsection{Polyominos and their dual graphs}
Let $T$ be the checkerboard tiling of $\R^2$ with unit squares, 
each square centered at a lattice point of $\Z^2$, and
where the square centered at the origin is white.
Let $W_0$ be the set of white squares both of whose coordinates
(the coordinates of the center of the square)
are even; let $W_1$ be the set of white squares both of whose coordinates are odd.
Let $B_0$ be the set of black squares whose coordinates are $(1,0)\bmod 2$
and $B_1$ the set of black squares whose coordinates are $(0,1)\bmod 2$.

A {\bf polyomino} is a finite\footnote{Later we will consider some special
infinite polyominos.} union of unit squares of $T$
bounded by disjoint simple closed lattice paths.
A corner of (the boundary of) 
a polyomino is {\bf convex} if the interior angle
is $\pi/2$; a corner is {\bf concave} if the interior angle is $3\pi/2$.
In either case the {\bf corner lattice square} is the lattice
square adjacent to the corner, which contains the 
angle bisector of interior angle.
An {\bf even polyomino} is a polyomino $P$ in which all corner squares
are of type $B_1$.
Note that this implies that 
any boundary edge of $P$ whose two corners are both convex or both concave
has odd length; any boundary edge
of $P$ with a convex and a concave corner has even length.
A polyomino is {\bf simply-connected} if it has only one boundary component.
\begin{lemma}\label{balanced}
A simply-connected
even polyomino contains one more black square than white square.
\end{lemma}
\begin{proof}
This is easily proved by induction on the number of corners,
starting from the case of a rectangle.
\end{proof}

\begin{figure}[htbp]
\begin{center}
\PSbox{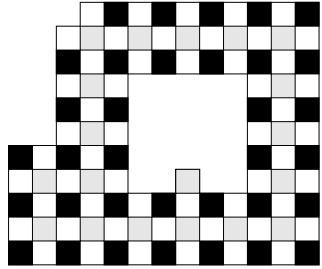}{1in}{2in}
\end{center}
\caption{\label{P}A Temperleyan polyomino. The black squares are in $B_1$,
the gray are in $B_0$.}
\end{figure}

A {\bf Temperleyan polyomino} 
is a polyomino which is obtained from an even polyomino
$P$ as follows. {\it Remove} from $P$ a black
lattice square $d_0$ adjacent to an edge or corner of the
outer boundary of $P$. For each interior boundary component $D_j$ of $P$,
{\it add} a black lattice square $d_j$ adjacent to an edge of that boundary.
We assume that $d_j$ only borders on a single square of $P$.
See Figure \ref{P}.
These added squares will be called {\bf exposed squares}.
Note that $d_0$ must be in $B_1$ and $d_j$ must be in $B_0$ for $j>0$.
From the lemma it follows that
a Temperleyan polyomino, even if not simply connected, contains the same number of black
squares as white squares.

Let $P$ be an even polyomino, and let
$\Bl(P)$ be the graph whose vertices are the squares $B_1$ in $P$, with edges connecting
all squares at distance $2$. Then to each horizontal edge of $\Bl(P)$ corresponds a square $W_1$ of $P$ (the square it crosses)
and to each vertical edge of $\Bl(P)$ corresponds a square of type $W_0$ of $P$.
To each face of $\Bl(P)$ which is not a boundary component of $P$ corresponds a square
of $P$ of type $B_0$. The planar graph $\Bl(P)$ has a planar dual $\Bo(P)$, whose vertices are
faces of $\Bo(P)$ (squares of type $B_0$), as well as a vertex for each boundary component of $P$.
For a Temperleyan polyomino constructed from $P$, we can still associate the same graphs $\Bl(P)$ and $\Bo(P)$,
but we mark the special vertex $d_0$ of $\Bl(P)$ and mark in $\Bo(P)$ the special edges adjacent to the $d_i$ for $i\geq 0$.

Temperley \cite{Temp} gave a bijection between spanning trees on an $m\times n$ grid
and domino tilings of a $(2m-1) \times (2n-1)$ polyomino with a corner removed.
A Temperleyan polyomino is a polyomino which arises from a subgraph of the grid by
a generalization of his construction, as above, where $\Bl(P)$ is the subgraph one starts with (see \cite{KPW}).

The {\bf interior dual graph} $M$ of a Temperleyan
polyomino $P$ is the graph with a vertex for each
lattice square in $P$, with edges joining pairs of vertices 
whose corresponding squares are at distance $1$ (in other words,
it is the dual graph without the boundary vertices).
Domino tilings of $P$ are in bijection with perfect
matchings of its interior dual graph (a perfect matching of a graph is a
set of edges such that each vertex is an endpoint of exactly one edge).
The exposed squares of $P$ are called {\bf exposed vertices} of $M$.

The interior dual graph $M$ of a polyomino $P$ is a subgraph of $\Z^2$
and its vertices inherit a coloring from the checkerboard coloring of 
the lattice squares:
$(x,y)$ is in $W_0$ if and only if $(x,y)\equiv(0,0)\bmod 2$ and so on.
We will usually denote a vertex $(x,y)\in\Z^2$ by the complex number $x+iy$.

\subsection{The height function}\label{ht}
\begin{figure}[htbp]
\begin{center}
\PSbox{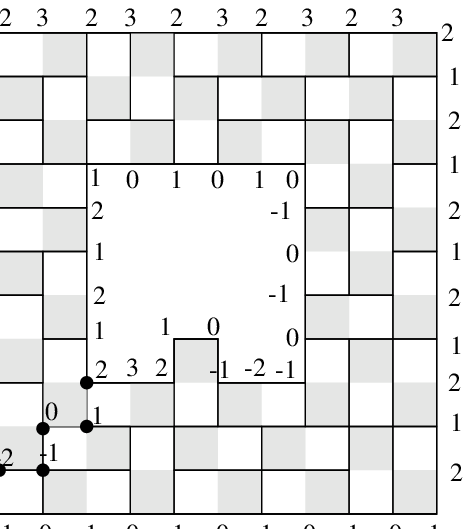}{1in}{2in}
\end{center}
\caption{\label{hts}Heights in a domino tiling.}
\end{figure}

Thurston \cite{Thurston} defines the height function on a domino tiling as
follows. 
The height function is a $\Z$-valued function on the vertices of the tiling,
defined only up to an additive constant.
Start at an arbitrary vertex of some domino and define the height there
to be $0$. For every other vertex $v$ in the tiling, 
take an edge-path $\gamma$ from $v_0$ to $v$
which follows the boundaries of the dominos. The height along $\gamma$
changes by $\pm1$ along each edge of $\gamma$: if the edge traversed
has a black square on its left (which may be exterior to the region)
then the height increases by $1$;
if it has a white square on its left then it decreases by $1$. This
defines a height at $v$. If the tiled region is simply connected,
the height is independent of the choice of $\gamma$
since the height change going around a domino is $0$.
If the tiled region is not simply connected the height is still
well-defined as long as each hole contains the same number of black and white
squares \cite{Thurston}.  See Figure \ref{hts}.

Let $M$ be the interior dual graph of a Temperleyan polyomino $P$,
and take a perfect matching of $M$. A height function on the tiling
determines a
height function defined
on the (non-boundary) faces of $M$.
The height function may be defined by assigning an arbitrary value
to some face and then applying 
the following rules: for each unmatched edge of $M$,
when following the edge from its black vertex to its white vertex,
the height of the face on the left minus the height of the face on
its right is $1$. For matched edges this difference is $-3$. 

\subsubsection{Heights of boundary components}
\label{bdyhts}
Let $P$ be a Temperleyan polyomino with boundary components $D_0,\ldots,D_k$
where $D_0$ is the outer component.
Since each $D_j$ encloses the same number of black squares as white squares
the net height change around each $D_j$ is zero, so the height
is well-defined for any tiling of $P$.

Given a tiling of $P$
the height function along $D_j$ depends only on the height
of any single point on $D_j$. That is, given two points $x_0,x_1$
of $D_j$, let $\gamma$ be the path running along $D_j$ from $x_0$ to $x_1$.
The height difference $h(x_1)-h(x_0)$ 
is independent of the tiling since $\gamma$
crosses no dominos.
Since the height of $D_j$ depends only on a single integer value,
it makes sense to talk about the height of $D_j$ as a single
$\Z$-valued random variable.

Note how the height changes as you go 
around a boundary component with the interior of $P$ on your left
(see Figure \ref{hts}). 
Along a straight edge the height alternates between two successive values.
Except at the exposed vertex, 
after a right turn the alternating pair decreases by $1$, and after a left
turn it increases by $1$ 
(this follows since all corners are black). This means that for two points on the
same boundary component, their height is related in a simple way to 
the amount of winding of the boundary component between them (i.e. the number of 
left turns minus the number of right turns). 

\subsection{Tilability of big Temperleyan polyominos}
The Temperleyan polyominos we will be using are those with small lattice spacing 
which approximate a region $U$ with smooth boundary (or piecewise smooth with
one-sided limits of tangents at each corner).
Tilability of such a polyomino can be shown using 
the following result of Fournier.
\begin{prop}[\cite{Fournier}] 
\label{fou}
A simply-connected polyomino
with the same number of black and white squares
can be domino-tiled unless there are two boundary vertices $x,y$ whose distance
in the $L^1$-metric (length of the shortest lattice path from $x$ to $y$
in $P$) is less than their height difference.
\end{prop}
Actually Fournier's condition
is stronger than this (he uses a modified metric)
but this will suffice for our needs.
Also, Fournier only considered simply-connected regions but his argument
generalizes to regions with many boundary components, as long as a height
has been assigned to each component (and one is interested in tilings
whose height function extends the function already defined on the boundary).

Since the region $U$ has a piecewise smooth boundary as defined above, the winding number of
the boundary
path between two points on the same boundary component of $U$ is bounded.
As a consequence if $P_\eps$ is a Temperleyan polyomino in $\eps\Z^2$ approximating
$U$ (and if locally the boundary of $P_\eps$ follows that of $U$ 
in the sense that
they are always directed into the same approximate quadrant), 
the height difference between two points on the same boundary 
component of $P_\eps$ is approximately 
the same as the winding number of the boundary of $U$ between those two points.
Therefore the height function on the boundary of $P_\eps$ 
varies by at most a constant.

In particular if $\epsilon$ is sufficiently small Proposition \ref{fou}
and Lemma \ref{balanced} show that $P_\eps$ is tilable.

A more elementary proof of tilability using spanning trees is sketched in
section \ref{trees}.

\section{Discrete analytic functions}\label{DAF}
The important discrete functions appearing in this article are examples 
of discrete analytic functions (also called monodiffric functions), see \cite{Duff}.  
This section reviews the relevant definitions.
Our definition is slightly different from the classical definition in
\cite{Duff} but is equivalent.

\subsection{The $\dbar$ operator}

We define several operators on $\Z^2$.
The operator $\partial_x\colon\C^{\Z^2}\to\C^{\Z^2}$ is defined
by: $$\partial_xf(v)=f(v+1)-f(v-1).$$
Similarly define
$$\partial_yf(v)=f(v+i)-f(v-i).$$
We define operators 
$$\partial_z=\partial_x-i\partial_y,$$
and
$$\dbar=\partial_x+i\partial_y.$$

These operators restrict to operators from $\C^B$ to $\C^W$: if $f\in\C^B$,
that is, if $f$ is zero on white vertices, then $\partial_xf,\partial_yf\in\C^W.$
Similarly $\partial_x,\partial_y$ map $\C^W$ to $\C^B$.
A {\bf discrete analytic function} is a function $F\in\C^B$ which is real
on $B_0$ and pure imaginary on $B_1$ and satisfies 
$\dbar F=0$.
If $F=f+ig$ where
$f\in\R^{B_0}$ and $g\in\R^{B_1}$, then
$F$ being discrete analytic is equivalent to $f$ and $g$ satisfying the 
{\bf discrete Cauchy-Riemann equations} 
\begin{eqnarray}
\label{CR1}
\partial_xf(v)&=&\partial_yg(v)~~\mbox{  for }v\in W_0\\
\label{CR2}
\partial_yf(v)&=&-\partial_xg(v)~~\mbox{  for }v\in W_1.
\end{eqnarray}
(Note that when $f\in\R^{B_0}$ and $g\in\R^{B_1}$, we have
$\partial_xf,\partial_yg\in\R^{W_0}$ and 
$\partial_yf,\partial_xg\in\R^{W_1}$.)

The function $f$ is called the real part of $f+ig$, and $g$
is called the imaginary part of $f+ig$.

If $f+ig$ satisfies the discrete CR-equations at all but
a finite number of (white) vertices, 
we say that $f+ig$ is discrete analytic with {\bf poles} at those vertices.

The operators $\partial_x,\partial_y,\partial_z,\dbar$
restrict to operators on subgraphs $M$ of $\Z^2$ in a natural
way: we consider $\C^M$ to be the subset of $\C^{\Z^2}$ which
consists of functions zero outside of $M$. We apply the operator
and then project back to $\C^M$.

\subsection{Laplacian}
A simple calculation shows that, if $f\in\R^{B_0}$, 
then $\partial_z \dbar f\in\R^{B_0}$ and $-\partial_z\dbar f$
is the Laplacian of $f$ on the graph $\Bo(\Z^2)$.
That is,
$$-\partial_z \dbar f(v)=\Delta f(v)=4f(v)-f(v+2)-f(v+2i)-f(v-2)-f(v-2i).$$

Note that this is $4$ times the usual Laplacian since we left out factors
of $\frac12$ in the definition of $\dbar$ and $\partial_z$.
Often when discussing the discrete Laplacian there is a disagreement
about the choice of sign. Here we chose the positive (semi-)definite
Laplacian, which corresponds in the continuous limit to 
$-\frac{\partial^2}{\partial x^2}-\frac{\partial^2}{\partial y^2}$.

In a similar fashion if $g\in\R^{B_1}$ then $-\partial_z \dbar g$ 
is the Laplacian of $g$ on the graph $\Bl(\Z^2)$.

In particular if $f+ig$ is discrete analytic on $\Z^2$ we have 
$\partial_z\dbar (f+ig)=\partial_z(0)=0$ and so
$\Delta f=0$ and $\Delta g=0$, where the first $\Delta$ is the Laplacian
on $\Bo(\Z^2)$ and the second is the Laplacian on $\Bl(\Z^2)$.

For a discussion of the boundary behavior of the Laplacian on $\Bo(P)$, see
section \ref{M'}.
\subsection{Weighting the graph}
\begin{figure}[htbp]
\begin{center}
\PSbox{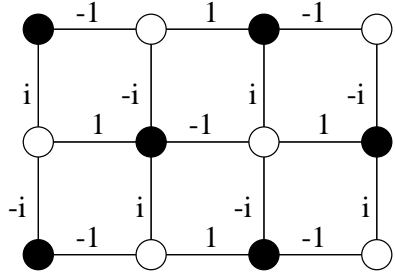}{1in}{2in}
\end{center}
\caption{\label{weights}Weights of the Kasteleyn matrix.}
\end{figure}

An alternative way to define discrete analytic functions,
which relates more closely with domino tilings, is as follows.
On the graph $\Z^2$ put {\bf weights} on the edges:
at each white vertex the four edge weights going counterclockwise
from the right-going edge are $1,i,-1,-i$ respectively.
See Fig. \ref{weights}.

Now for a pair of real-valued functions $f\in \R^{B_0}$
and $g\in\R^{B_1},$ the function $f+ig$ is discrete analytic 
if and only if 
it satisfies $K(f+ig)=0$, where $K$ is the adjacency matrix
of $\Z^2$ with these weights. The matrix $K$ is called 
the {\bf Kasteleyn matrix} of $\Z^2$.  Kasteleyn proved that for a finite
region the absolute value of the determinant
of the Kasteleyn matrix is the square of the number of
perfect matchings. (Usually the Kasteleyn matrix is defined with different weights
\cite{Kast};
but in fact any choice of complex weights of modulus $1$ satisfying $ac=-bd$ for the four weights
$a,b,c,d$ around a square gives rise to a Kasteleyn-like matrix
whose determinant counts tilings.)

When considered as an operator on $\C^B$,
the operator $K$ 
is the operator $\dbar$. When considered
as an operator on $\C^W,$ however, it is $-\dbar$.
Let $K^*$ be the Hermitian conjugate of $K$.
Then the operator $K^*K$ is acting as the Laplacian on both
$\Bo$ and $\Bl$.

\begin{lemma}
A discrete analytic function on a simply connected
Temperleyan region $P$ is determined up to an additive
(imaginary) constant by its real part.
\end{lemma}
\begin{proof}
Note first that $\Bl(P)$ is connected.
Let $f\in \R^{B_0}$ be harmonic on $\Bo(P)$.
Given the value of the imaginary
part $g$ at one vertex $v\in B_1$, the value $g(w)$ for any other vertex $w$
in $B_1$ is uniquely determined as follows. Take a path in $\Bl(P)$
from $v$ to $w$. Each edge of the path crosses an edge of $\Bo(P)$. 
One of the Cauchy-Riemann equations ((\ref{CR1}) or (\ref{CR2}))
at the crossing point determines
the difference in values of $g$ at the endpoints of this edge.
The value $g(w)$ is obtained by summing this difference along the
path.
The harmonicity of $f$ implies that the value $g(w)$ obtained
is independent of the path chosen. 
\end{proof}

When the region is not simply connected, in general the conjugate function
of a harmonic function $f\in\R^{B_0}$
is not single-valued: the ``integral'' in the above lemma along a path surrounding a hole may not be zero.

\section{The coupling function}\label{boundary}
Let $M$ be the interior dual graph of a Temperleyan polyomino $P$. Let 
$K$ be the corresponding Kasteleyn matrix 
and let $E$ be a finite collection of disjoint edges of $M$.
Let $b_1,\ldots,b_k$ and $w_1,\ldots,w_k$ be the black vertices
(respectively white vertices) covered by $E$.
Let $\mu$ be the uniform probability measure on
perfect matchings of $M$.
\begin{thm}[\cite{Kenyon}]\label{ken}
The $\mu$-probability that $E$ occurs in a perfect matching
is given by $|\det(K^{-1}_E)|$, where $K^{-1}_E$ is the 
submatrix of $K^{-1}$ whose rows are indexed by $b_1,\ldots,b_k$
and columns are indexed by $w_1,\ldots,w_k$.
More precisely, the probability is
$(-1)^{\sum p_i+q_i}a_E \det(K^{-1}_E)c$,
where $p_i,q_i$ is the index of $b_i$, resp. $w_i$, in a fixed ordering
of the vertices, $c=\pm 1$ is a constant depending only on that ordering,
and $a_E$ is the product of the edge weights of the edges $E$.
\end{thm}

Thus the $\mu$-measures of cylinder sets for perfect matchings on $M$
are determined by this function
$K^{-1}\colon M\times M\to\C$, called the {\bf coupling function}.
For historical reasons we denote the coupling function with a $C$.

Actually this theorem holds for arbitrary bipartite planar graphs, not just
those arising from the square grid: see \cite{Kenyon}.

In all of our applications of this theorem we will use only a small
number of edges out of the total number of edges of $M$; in this case
we can choose the ordering of vertices so that all the relevant
indices $p_i$ and $q_i$ are even, and $c=1$. Then we can use the simpler form
$|\det(K^{-1}_E)|=a_E\det(K^{-1}_E)$.

The defining property of $C(v_1,v_2)$ is that it satisfies:
$KC(v_1,v_2)=\delta_{v_1}(v_2)$. Here $\delta_{v_1}$ is the delta function
$$\delta_{v_1}(v_2)=\left\{\begin{array}{ll}1&\mbox{if }v_2=v_1\\
0&\mbox{otherwise.}\end{array}\right.$$
We have the following.

\begin{lemma}\label{8}
The function $C$ is symmetric: $C(v_1,v_2)=C(v_2,v_1)$.
We have $C(v_1,v_2)=0$ whenever $v_1$ and $v_2$ are both black or both white.
If $v_1$ is white,
the coupling function $C(v_1,v_2)$ is discrete analytic as a function
of $v_2$, with a pole at $v_1$. 
\end{lemma}

\begin{proof} 
Since we already have $KC(v_1,v_2)=\delta_{v_1}(v_2)$, it suffices
to show that $C(v_1,v_2)$ is real when $v_2-v_1\equiv(1,0)\bmod 2$,
pure imaginary when $v_2-v_1\equiv(0,1)\bmod 2$ and zero in the remaining cases.

If we order the vertices of $M$ in such a way that all the $W_0$ are first,
then $W_1$ then $B_0$ and then $B_1$, then the matrix $K$ in this basis
has the form
$$K=\left(\begin{array}{cccc}0&0&K_1&iK_2\\0&0&iK_3&K_4\\K_1^t&iK_3^t&0&0\\
iK_2^t&K_4^t&0&0\end{array}\right)$$
where $K_1,K_2,K_3,K_4$ are real matrices. The conjugate of the above matrix
by the matrix 
$$\left(\begin{array}{cccc}I&0&0&0\\0&iI&0&0\\0&0&I&0\\0&0&0&iI
\end{array}\right)$$
is real. Hence the 
inverse of $K$ has the same form as $K$.
This completes the proof.
\end{proof}

See Figure \ref{couplingfn} for (part of) an example.

Since $C(v_1,v_2)=0$ when $v_1,v_2$ are both black or both white,
and $C(v_1,v_2)=C(v_1,v_2),$ we will almost always take the first argument of $C$ to be
a white vertex and the second to be black.

\subsection{Boundary conditions for the coupling function}
\label{M'}
A discrete analytic function is determined by its boundary values,
since its real and imaginary parts are harmonic.
In this section we describe the behavior of $C(v_1,v_2)$ for $v_2$
on the boundary of $M$. 

\begin{figure}[htbp]
\begin{center}
\PSbox{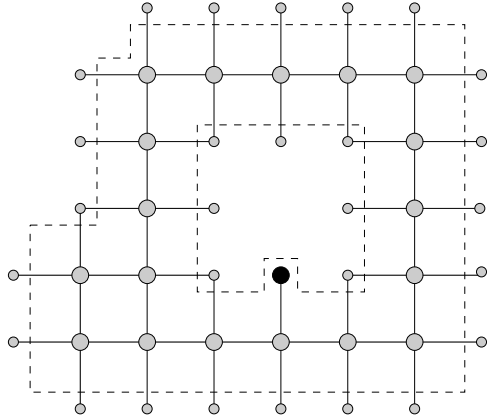}{1in}{2.5in}
\end{center}
\caption{\label{graphex}Example of the graph $\Bo'(P)$ for the polyomino $P$
of Figure \protect\ref{P} ($P$ is in dashed lines). 
The smaller gray dots are vertices
in $Y$; the black vertex is the exposed vertex.}
\end{figure}

Assume that $v_1\in W_0$. 
By Lemma \ref{8}, $C(v_1,v_2)$ is real when $v_2\in \Bo(P)$
and pure imaginary when $v_2\in\Bl(P)$ (and zero when $v_2\in W_0\cup W_1$).
Let $Y$ be the set of vertices in $B_0$ adjacent to (a white vertex of) $M$ but not in $M$
(that is, at distance $1$ from a vertex of $M$).
Let $\Bo'(P)$ be the graph whose vertices are
$\Bo(P)\cup Y$, and whose edges connect every pair of vertices of distance $2$,
provided that the white vertex lying between these two is in $M$.
The set $Y$ is the set of {\bf boundary vertices of $\Bo'(P)$}.
Let $V$ be the set of exposed vertices $d_1,\ldots,d_k$ (recall that
they are all in $B_0$).
See Fig. \ref{graphex} for an example of a graph $\Bo'(P)$.

\begin{lemma}\label{smallcase}
For a fixed $v_1\in W_0$, consider $C(v_1,v_2)$ as a function of $v_2$.
The real part of $C(v_1,v_2)$, extended to be zero on $Y$ and
considered as a function on the graph $\Bo'(P)$,
has the following properties:
\begin{enumerate}
\item it is harmonic at all vertices in $\Bo(P)\setminus(V\cup\{v_1+1,v_1-1\})$.
\item $\Delta\Re C(v_1,v_1\pm1)=\pm1$,
\item its harmonic conjugate is single-valued.
\end{enumerate}
If rather $v_1\in W_1$ then the imaginary part of $C(v_1,v_2)$, extended
to be zero on $Y$ and considered as a function on $\Bo'(P)$, has
the following properties:
\begin{enumerate}
\item it is harmonic at all vertices in $\Bo(P)\setminus(V\cup\{v_1+i,v_1-i\})$.
\item $\Delta\Im C(v_1,v_1\pm i)=\mp1$,
\item its harmonic conjugate is single-valued.
\end{enumerate}
\end{lemma}

\begin{proof}
The first two properties in both cases follow from 
$$\Delta C(v_1,\cdot)=K^*K C(v_1,\cdot)=
K^*\delta_{v_1}=
\delta_{v_1+1}-\delta_{v_1-1}-i\delta_{v_1+i}+i\delta_{v_1-i}.$$
This equation is valid at every vertex of $\Bo(P)$ except the exposed
vertices (which do not have $4$ neighbors).
The third property in each case follows by definition, since $\Im C(v_1,\cdot)$
is the harmonic conjuate of $\Re C(v_1,\cdot)$ and $-\Re C(v_1,\cdot)$
is the harmonic conjugate of $\Im C(v_1,\cdot)$.
\end{proof}

We will see later that 
$\Re C(v_1,v_2),\Im C(v_1,v_2)$ are respectively
the unique functions with the above properties.
As a consequence we will be able to use some general
theorems about harmonic 
functions to reach conclusions about the coupling function.

The conditions in Lemma \ref{smallcase} are particularly simple 
because we started with a Temperleyan polyomino. For a polyomino
with different boundary conditions, the corresponding boundary conditions
for the coupling function can be quite complicated: see 
section \ref{conclusion}.

\section{Asymptotic values of the coupling function}\label{limits}
Here we will show that, as $\epsilon$ tends to $0$, the 
scaled discrete analytic function $\frac1\epsilon C(v_1,\cdot)$ 
converges to a pair of complex-analytic functions $F_0,F_1$ 
($F_0$ when $v_1\in W_0$ and $F_1$ when $v_1\in W_1$) which
transform analytically (see Proposition \ref{Finvariant})
under conformal mappings of the domain $U$.

We first study what happens when the polyomino
$P$ is the whole plane, since as we will see,
for any region $U$ the leading term in $C(v_1,v_2)$ equals $C_0(v_1,v_2)$, 
the coupling function on the plane
(as long as $v_1$ is not too close to the boundary of $U$).

\subsection{On the plane}
In \cite{Kenyon} we gave an explicit formula for the
coupling function on $\Z^2$. 
This was shown to be the limit as $n\to\infty$ of the coupling function on
the $2n\times 2n$ square, centered at the origin.
In that paper we used different weights for the Kasteleyn
matrix:
$1$ on all horizontal edges and $i$ on all vertical edges.
The present calculation is straightforward using the same methods
(in fact the result is identical after changing the sign on alternating
vertices of $B_0$ and $B_1$) and yields the following.
\begin{prop}[\cite{Kenyon}]\label{C0form}
Let $C_0$ denote the coupling function for the whole plane $\Z^2$. Then
$$C_0(0,x+iy)=\frac{1}{4\pi^2}\int_0^{2\pi}\int_0^{2\pi}\frac{
e^{i(x\theta-y\phi)}}{2i\sin(\theta)+2\sin(\phi)}d\theta d\phi.$$
\end{prop}

\begin{figure}[htbp]
\begin{center}
\PSbox{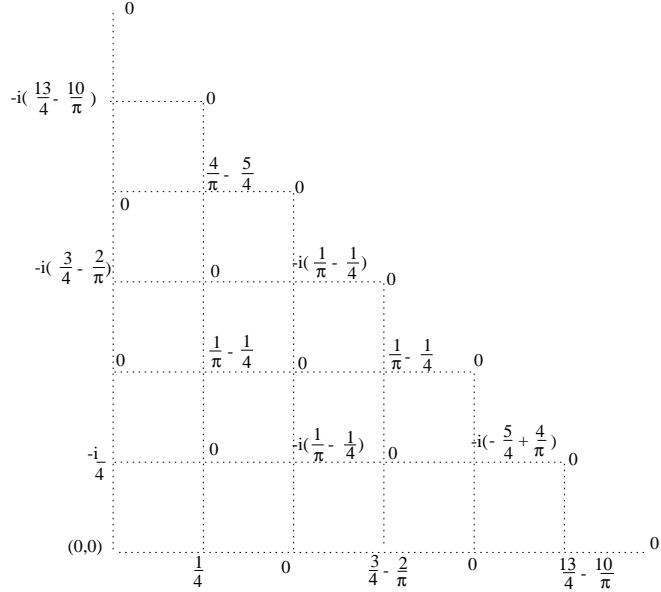}{1in}{3in}
\end{center}
\caption{\label{couplingfn}The function $C_0(0,x+iy)$, 
the coupling function for $\Z^2$.}
\end{figure}

By translation invariance, $C_0(v_1,v_2)=C_0(0,v_2-v_1)$ so this theorem
describes the entire coupling function.
In \cite{Kenyon} it is shown how to evaluate explicitly this integral.
Figure \ref{couplingfn} shows the first few values of $C_0(0,x+iy)$
when $x+iy$ is in the positive quadrant. The values in the other
quadrants are obtained by the symmetry $C_0(0,iz)=-iC_0(0,z)$,
which arises from the corresponding symmetry of the edge weights.

Recall that the origin in $\Z^2$ is a vertex of type $W_0$.
\begin{thm}
\label{1/z}
As $|z|\to\infty$, the coupling function on $\Z^2$
is asymptotically equal to $\frac1{\pi z}$, that is
$$C_0(0,z)=\left\{\begin{array}{ll}
\mbox{Re}\frac1{\pi z}+O(\frac1{|z|^2}) &z\in B_0\\
i\mbox{Im}\frac1{\pi z}+O(\frac1{|z|^2}) &z\in B_1.\end{array}\right.
$$
\end{thm}

\begin{proof}
There is the following relation between $C_0$ and the Green's function for
the plane. The real part of $C_0$ is the unique function 
on $\Bo(\Z^2)$ 
satisfying $\Delta Re C_0=\delta_1-\delta_{-1}$ and tending to $0$ at infinity
(see Lemma \ref{smallcase}, and recall
that $C_0$ is the limit of $C$ on square regions centered at the origin).

Now the classical Green's function $G_0(v,w)$ on $\Z^2$ satisfies
$\Delta G_0(0,w)=\delta_0(w)$ and for any fixed $v$,
$G_0(0,w)-G_0(v,w)\to 0$ as $w\to\infty$
(see Lemma \ref{G0asymp}).
As a consequence we have
$$\Re C_0(0,w)=G_0(0,\frac{w-1}2)-G_0(0,\frac{w+1}2),$$
where on the right we used coordinates on $\Bo(\Z^2)$ which has 
index $4$ in $\Z^2$.

Using Lemma \ref{G0asymp} we have
\begin{eqnarray*}
\Re C_0(0,w)&=&G_0(0,\frac{w-1}2)-G_0(0,\frac{w+1}2)\\
&=&\frac1{2\pi}\left(\log|\frac{w+1}2|-\log|\frac{w-1}2|\right)+O(\frac1{|w|^2})\\
&=&\frac1{2\pi}\Re\log(\frac{w+1}{w-1})+O(\frac1{|w|^2})\\
&=&\frac1{2\pi}\Re\frac2{w-1}+O(\frac1{|w|^2})\\
&=&\Re\frac1{\pi w}+O(\frac1{|w|^2})
\end{eqnarray*}
where we used $\log(1+z)=z+O(|z|^2)$.
A similar argument holds for the imaginary part.
\end{proof}

\begin{lemma}[\cite{Stohr}]\label{G0asymp}
For the Green's function $G_0$ on $\Bo(\Z^2)$ we have
\begin{equation}\label{g0as}
G_0(0,v)=-\frac1{2\pi}\log|v| + c_0 + O(\frac1{|v|^2})
\end{equation}
for a constant $c_0$.
\end{lemma}
Note that St\"ohr's Laplacian is $-1/4$ times ours, so his Green's
function is $-4$ times that in (\ref{g0as}).

\subsection{The half-plane}
For later use we will need to compute the coupling function on
a half-plane.
Let $\{P_n\}$ be a sequence of Temperleyan polyominos in the upper half
plane $H=\{x+iy\in\Z^2~|~y>0\}$, such that
$P_n$ contains the rectangle $[-n,n]\times[1,n]$, and the base point
$d_0$ of $P_n$ is outside this rectangle. Then (as we will show in 
the proof of Theorem
\ref{nearbdy}), for fixed $v_1,v_2$
the coupling function $C^{(n)}(v_1,v_2)$ on $P_n$ converges
to a limit $C_H(v_1,v_2)$ satisfying the properties below.
In particular the uniform measures on the $P^{(n)}$ converge to a unique
measure $\mu_H$.

Suppose $v_1\in W_0$.
The real part of $C_H(v_1,v_2)$ satisfies the conditions of Lemma 
\ref{smallcase}:
$\Delta\mbox{Re}C_H(v_1,\cdot)=\delta_{v_1+1}-\delta_{v_1-1}$,
$\mbox{Re}C_H(v_1,x+iy)=0$ when $y=0$, and $\mbox{Re}C_H$ 
tends to zero at infinity. There is
a unique harmonic function with these three properties: the real part of
$C_0(v_1,v_2)-C_0(\overline{v_1},v_2)$ (note that $v_1\in W_0$ implies
$\overline{v_1}\in W_0$). The conjugate harmonic function $\Im C_H$
is single-valued, and uniquely defined by the condition that it tends to zero at
infinity; as a consequence we have
\begin{equation}\label{polescancel}
C_H(v_1,v_2)=C_0(v_1,v_2)-C_0(\overline{v_1},v_2)\mbox{   when }v_1\in W_0.
\end{equation}

If $v_1\in W_1$, on the other hand, it is the {\it imaginary part}
of $C_H(v_1,x+iy)$ which is zero when $y=0$. In this case there is again a
unique harmonic function satisfying the requisite properties:
$\mbox{Re}C_H(v_1,v_2)=\mbox{Re}(C_0(v_1,v_2)+C_0(\overline{v_1},v_2))$.
So then
\begin{equation}
\label{polesdontcancel}
C_H(v_1,v_2)=C_0(v_1,v_2)+C_0(\overline{v_1},v_2)\mbox{   when }v_1\in W_1.
\end{equation}

There is a big difference between these two cases: from Theorem \ref{1/z},
in the case $v_1\in W_0$ we have
$$C_H(v_1,v_2)=\frac1\pi\left(\frac1{v_2-v_1}-\frac1{v_2-\overline{v_1}}\right)+
O(\frac1{|v_2-v_1|^2})$$
$$=\frac{v_1-\overline{v_1}}{\pi(v_2-v_1)(v_2-\overline{v_1})}+
O(\frac1{|v_2-v_1|^2})$$
which is $O(d)$, where $d$ is the distance from $v_1$ to the boundary.
In the case $v_1\in W_1$, rather, we have
$$C_H(v_1,v_2)=\frac1\pi\left(\frac1{v_2-v_1}+\frac1{v_2-\overline{v_1}}\right)+
O(\frac1{|v_2-v_1|^2})$$
$$=\frac{2v_2-v_1-\overline{v_1}}{\pi(v_2-v_1)(v_2-\overline{v_1})}+
O(\frac1{|v_2-v_1|^2})$$
which does not go to zero as $v_1$ approaches the boundary.

There are similar formulas for the other half-planes with horizontal
or vertical boundary.

\subsection{Bounded regions}
\label{approx}
One of the main results in this paper is to show that the coupling function
on a finite region converges,
as $\eps$ tends to zero, to a pair of analytic functions
which transform analytically under conformal maps of the region.
For a fixed region $U$ we can not prove this for all 
Temperleyan polyominos $P_\eps$ approximating $U$:
we require that the approximating $P_\eps$ have a nice behavior in a neighborhood of their
exposed vertices. This shortcoming is due to our lack of understanding
of the asymptotics of the discrete Green's function near the boundary of a polyomino.
It seems nonetheless reasonable to suspect that this flaw can and will be overcome in the near future.

We will begin at this point to 
use the metric on $\eps\Z^2$ rather than $\Z^2$.
That is, we work on polyominos in $\eps\Z^2$ with interior dual graphs
having edges of length $\eps$. The graphs $\Bo'(P)$ have edges of length $2\eps$.

Let $U$ be a region in $\C$ with smooth boundary (or piecewise smooth as previously defined).
Let $D_0,\ldots,D_k$ be the boundary components
of $U$, with $D_0$ being the outer component. Let $d_j'$ be a marked
point of $D_j$. 
Let $z_1$ be a point in the interior of $U$ and $z_2$ be any point of $U$.

We define two functions $F_0(z_1,z_2)$ and $F_1(z_1,z_2)$,
whose existence and uniqueness will be shown in the proof of Theorem
\ref{longrange}, below.
For fixed $z_1$, 
the function $F_0(z_1,z_2)$ is analytic as a function of $z_2$, has a simple pole
of residue $1/\pi$ at $z_2=z_1$ and no other poles on $\overline U$ except
possibly simple poles at the $d_j',~j>0$. Furthermore it is zero at $d_0'$ 
and has real part $0$ on the boundary of $U$.
For fixed $z_1$, the 
function $F_1(z_1,z_2)$ is analytic as a function of $z_2$, has a simple
pole
of residue $1/\pi$ at $z_2=z_1$ and no other poles on $\overline U$ except
possibly simple poles at the $d_j',~j>0$. Furthermore it is zero at $d_0'$
and has imaginary part $0$ on the boundary of $U$.

For each $\epsilon>0$ sufficiently small,
let $P_\eps$ be a Temperleyan polyomino in $\epsilon\Z^2$ approximating 
$U$ in the following sense. 
The boundaries of $P_\eps$ are within $O(\eps)$ of the boundaries of $U$, 
and except near a corner of $\partial U$ the tangent vector to $\partial U$ points
into the same halfspace as the direction of the corresponding edges of $
\partial P_\eps$.
Furthermore assume that the exposed vertices $d_j$ of $P_\eps$
are within $O(\eps)$ of the $d_j'$.
Suppose further that for a certain $\delta=\delta(\eps)>0$ 
tending to zero sufficiently slowly (see below), 
in a $\delta$-neighborhood of each $d_j$,
the boundary of $P_\eps$ is straight (horizontal or vertical).
Let $M_\eps$ be the interior dual of $P_\eps$.
Let $v_1$ be a white vertex and $v_2$
a black vertex of $M_\eps$.
We then have the following result.
\begin{thm}
\label{longrange} Fix any real $\xi>0$.
The coupling function $C(v_1,v_2)$
on the graph $M_\eps$ satisfies:
for $v_1\in W_0$ and 
$v_1,v_2$ not within $\xi$ of the boundary of $M_\eps$,
$$\frac1{\epsilon}C(v_1,v_2)=\frac1\eps C_0(v_1,v_2)+F_0^*(v_1,v_2)+o(1),$$
where $F_0^*$ is defined by the condition that $F_0(z_1,z_2)=
\frac1{\pi(z_2-z_1)} + F_0^*(z_1,z_2),$ with $F_0$ as above, and $C_0$
is the coupling function on $\eps\Z^2$.

If $v_1\in W_1$, rather, then 
$$\frac1{\epsilon}C(v_1,v_2)=\frac1\eps C_0(v_1,v_2)+F_1^*(v_1,v_2)+o(1),$$
where $F_1^*$ is defined by the condition that $F_1(z_1,z_2)=
\frac1{\pi(z_2-z_1)} + F_1^*(z_1,z_2)$, with $F_1$ as above.
\end{thm}

The equality in the theorem should be interpreted as saying:
when $v_1\in W_0$ and 
$v_2\in B_0$ then $C(v_1,v_2)$ equals the real part of the right-hand
side; and when $v_1\in W_0$ and 
$v_2\in B_1$ then $C(v_1,v_2)$ equals $i$ times the imaginary part
of the right-hand side. Similarly for $v_1\in W_1$ and $v_2\in B_0$,
then $C(v_1,v_2)$ equals $i$ times the imaginary part of the right-hand
side; when $v_1\in W_1$ and 
$v_2\in B_1$ then $C(v_1,v_2)$ equals the real part of the right-hand side.

When $v_1$ and $v_2$ are far apart (not within $o(1)$)
then we can replace
$\frac1\eps C_0(v_1,v_2)$ with $\frac1{\pi(v_2-v_1)}+o(1)$ and so
the statement is simply
$$\frac1\eps C(v_1,v_2)=F_j(v_1,v_2)+o(1)$$ where $j=0$ or $1$
as the case may be.

\begin{proof} Let $U_\delta$ be equal to $U$ except in a $2\delta$-neighborhood 
of the $d_j'$,
and such that $U_\delta$ is flat and horizontal or vertical
in a $\delta$-neighborhood of the $d_j'$.
We will first prove the theorem for $U_\delta$ for any {\it fixed} $\delta>0$.

We will do only the case $v_1\in W_0$. The case $v_1\in W_1$
is identical using the imaginary part of $C$ rather than the real part of $C$
below.

Let $G(w_1,w_2)$ be the Green's function on $\Bo'(P_\eps)$ (recall
the construction of $\Bo'(P_\eps)$ from section \ref{M'}), that is,
the function which satisfies $\Delta G(w_1,w_2)=\delta_{w_1}(w_2)$
and $G(w_1,w_2)=0$ when $w_2\in Y\cup V\setminus\{w_1\}$.

The function $\Re C(v_1,v_2)$, considered as a function of $v_2$,
is a linear combination
of the Green's functions $G(v_1\pm\eps,v_2)$ and $G(d_j,v_2)$ for $j=1,\dots,k$
since it is harmonic off of these
vertices. In fact since 
$$\Delta\Re C(v_1,\cdot)=\delta_{v_1+\eps}-\delta_{v_1-\eps}+
\sum_{j=1}^k\alpha_j\delta_{d_j}$$
for some constants $\alpha_j$, we have
\begin{equation}\label{potl}
\Re C(v_1,v_2)=G(v_1+\eps,v_2)-G(v_1-\eps,v_2)+\sum_{j=1}^k \alpha_j G(d_j,v_2).
\end{equation}

By Corollary \ref{bdygreen} 
below, the rescaled Green's function $\frac1\eps G(d_j,v_2)$
(considered as a function of $v_2$) converges away from $d_j$
to a continuous harmonic function with a logarithmic singularity at $d_j$ and 
boundary values $0$. 
(This is the place where we need $U_\delta$ rather than $U$.)
Similarly by Lemma \ref{GGconv} the difference $\frac1\eps(
G(v_1+\eps,v_2)-G(v_1-\eps,v_2))$ converges.
It remains to show that the coefficients $\alpha_j$
in (\ref{potl})
converge as $\epsilon\to 0$.
Note that if $U$ is simply connected then $k=0$ and we are done.

For general $U$, the right hand side of (\ref{potl}) 
automatically satisfies the conditions (1) and (2) of Lemma \ref{smallcase}
defining the coupling function,
but the Green's functions $G(v_1,v_2)$ do not in general have single-valued
harmonic conjugate. It is necessary to choose the $\alpha_j$ so
that the harmonic conjugate of the right-hand side of (\ref{potl}) is 
single-valued.  We show that in fact
the $\alpha_j$ are {\it uniquely determined} by this property.

We will use the language of electrical networks, see e.g.\! \cite{DS}. 
Consider the graph $\Bo'(P)_\eps$ 
to be a resistor network with resistances $1$ on each edge.
The function $G(v_1,v_2)$ is the potential
at $v_2$ when one unit of current flows into the network at $v_1$ and the
boundary $Y\cup V$ is held at potential $0$.
The $\alpha_j$ must be chosen so that, when 
currents $\alpha_j$ flow into the network at $d_j$, and current $\pm 1$
flows into the network at $v_1\pm\eps$, and the boundary is held at potential
$0$, then the net amount of current exiting
each boundary component $D_j$ is zero. For, the harmonic conjugate 
is the integral of the current flow: the integral of the current
crossing a closed curve surrounding $D_j$ is $0$ if and only if the harmonic
conjugate is single-valued around that curve.

We claim that given any $k+1$ real numbers $c_0,c_1,\ldots,c_k$
such that $c_0+\dots+c_k=0$, there exists a unique choice of 
reals $\alpha_1,\dots,\alpha_k$ such that, when currents $\alpha_j$ 
flow into the network at $d_j$,
and the boundary is held at potential $0$,
the net current flow out of each boundary component $D_j$ is $c_j$.
This will then determine the $\alpha_j$, because letting 
$c_0,\dots,c_k$ be the current
flow out of the boundaries
from the function $G(v_1+\eps,v_2)-G(v_1-\eps,v_2)$ (we mean,
when $1$ unit of current flows in at $v_1+\eps$ and $1$ flows out at $v_1-\eps$), 
we must choose the unique $\alpha_j$ to exactly cancel this flow.

To prove the claim, note that the map $\Phi\colon\R^k\to\R^k$
which gives the outgoing currents $c_1,\dots,c_k$ (and therefore 
$c_0=-c_1-\dots-c_k$ as well) as a function of $\alpha_1,\dots,\alpha_k$
is linear (this is the principle of superposition). 
It suffices to show that the determinant of $\Phi$ is nonzero. 

However on each column of the matrix of $\Phi$ (in the basis
$\{c_1,\dots,c_k\}$ and $\{\alpha_1,\dots,\alpha_k\}$)
the diagonal entry is the only 
negative entry: $G(d_j,v_2)$ induces a positive net current flow out of each
boundary component except the component $D_j$ which contains $d_j$, 
since $G(d_j,v_2)$ is a {\it positive}
harmonic function.  Furthermore the diagonal entry in $\Phi$ is larger than
the absolute value of the sum of the other entries in that column,
since a nonzero amount of current flows out of $D_0$, i.e. $c_0>0$
(and the total inflowing current equals the total outflowing current).
This implies that $\det\Phi\neq 0$ (see Lemma \ref{detneq0} below).

Now as $\eps$ tends to $0$, the rescaled Green's function
$\frac1\eps G(d_j,v_2)$ converges (Corollary \ref{bdygreen}). 
This implies that the entries of the matrix of
$\Phi$ converge: the pointwise convergence 
of a sequence of harmonic functions implies convergence
of their derivatives (even in the discrete case), due to Poisson's formula:
the derivative at a point
is determined by integrating the values of the function on a neighborhood
of that point against (the derivative of) the Poisson kernel. By integrating the derivative
we get convergence of the net current flow out of each boundary.
Furthermore the amount of current out of $D_0$
due to $\frac1\eps G(d_j,\cdot)$
is bounded from below. This implies that
$\det\Phi$ is bounded away from $0$ (Lemma \ref{detneq0}). Since the 
difference in Green's 
functions $\frac1\eps G(v_1+\eps,\cdot)-\frac1\eps G(v_1-\eps,\cdot)$ 
also converges (Lemma \ref{GGconv}), the net current out of $D_j$
from $\frac1\eps G(v_1+\eps,\cdot)-\frac1\eps G(v_1-\eps,\cdot)$ converges. 
Therefore the $\alpha_j$ converge as well.
We conclude that $\Re C$ converges.

The $C^0$-convergence of $\Re C$ implies convergence
of its derivatives and so by integrating we get local
convergence of $\Im C$ as well.
By uniqueness of the harmonic conjugate (up to an additive constant)
we have that $\Im C$ converges 
(the constant is determined by the fact that it is zero at $d_0$).

In conclusion when $v_1\in W_0$, 
$\frac1\eps C(v_1,v_2)$ converges to an analytic function (of $v_2$)
with all the properties of the function $F_0$. Furthermore the proof shows
that there is a unique function with these properties. 
When $v_1\in W_1$ then $C(v_1,v_2)$ converges to $F_1$ which is also
unique.

When $|v_2-v_1|=o(1)$, the main
contribution to $C(v_1,v_2)$ is from $G(v_1+\eps,v_2)-G(v_1-\eps,v_2)$;
the unrescaled Green's functions $\alpha_jG(d_j,v_2)$ contribute at most $o(1)$.
Since $G(v_1+\eps,v_2)-G(v_1-\eps,v_2)=G_0(v_1+\eps,v_2)-G_0(v_1-\eps,v_2)+o(1)$
(see the proof of Lemma \ref{GGconv}),
we conclude that $C(v_1,v_2)=C_0(v_1,v_2)+o(1)$. This gives the ``local''
term in the statement.

The above holds for $U_\delta$ for any $\delta>0$. It remains to see that
when $\delta\to0$
the functions $F_0^{(\delta)},F_1^{(\delta)}$ on $U_\delta$ converge to $F_0,F_1$
on $U$. This follows from Proposition \ref{Finvariant} below, and the fact that
the Riemann map from $U_\delta$ to $U$ converges (if appropriately normalized)
to the identity mapping.
Therefore the result holds for $U$ as long as $\delta\to0$
sufficiently slowly.
\end{proof}

A similar result holds when $v_1$ is close to a flat boundary of $P_\eps$.
Here is the statement when it is close to a flat horizontal boundary.
This is the only case we will need later.
\begin{thm}\label{nearbdy} Fix $\delta>0$.
Let $z_1$ be a point on the boundary of $U$ such that the boundary is flat 
and horizontal in a $\delta$-neighborhood of $z_1$. 
Let $v_1\in W_0$ be a point within $O(\eps)$ of $z_1$ and $v_2$ a black vertex.
The coupling function $C(v_1,v_2)$ satisfies
$$\frac1\eps C(v_1,v_2)=\frac1\eps 
C_H(v_1,v_2)+o(1)$$
where $C_H$ is the coupling function defined in (\ref{polescancel})
for appropriate half-plane in $\eps\Z^2$.
If rather $v_1\in W_1$ then
$$\frac1\eps C(v_1,v_2)=\frac1\eps C_H(v_1,v_2)+
F_1^{**}(z_1,v_2)+o(1),$$
where $F_1^{**}$ is defined by the condition that $F_1(z_1,z_2)=
\frac2{\pi(z_2-z_1)} + F_1^{**}(z_1,z_2)$ and $F_1$ is as before.
\end{thm}

\begin{proof} We use the notation of the previous proof.
If $v_1\in W_0$, then by (\ref{polescancel}),
the function $\frac1\eps(G_0(v_1+\eps,v_2)-G_0(v_1-\eps,v_2))$ is already $o(1)$
for $v_2$ near the boundary of $U$ except at the point $z_1$. 
Therefore the $\alpha_j$ will all tend to $0$ as well. The result 
follows if we define $C_H(v_1,v_2)=G_0(v_1+\eps,v_2)-G_0(v_1-\eps,v_2)$.

On the other hand if $v_1\in W_1$, then by (\ref{polesdontcancel}),
the function $\frac1\eps(G_0(v_1+i\eps,v_2)+G_0(v_1-i\eps,v_2))$ 
has two poles (each of residue $1/\pi$)
within $o(1)$ of $v_1$. The remainder of the proof is similar to that of the
previous theorem.
\end{proof}

Again note that when $v_2$ and $v_1$ are not close, in case $v_1\in W_0$ we have
$\frac1\eps C(v_1,v_2)=F_0(v_1,v_2)+o(1)=o(1)$ and
when $v_1\in W_1$ we have 
$\frac1\eps C(v_1,v_2)=F_1(v_1,v_2)+o(1).$

The functions $F_0,F_1$ 
depend only on the conformal type of the domain $U$ in the following sense.
Let $F_+=F_0+F_1$ and $F_-=F_0-F_1$.

\begin{prop}\label{Finvariant} The function $F_+(z_1,z_2)$
is analytic as a function of both variables. The function
$F_-(z_1,z_2)$ is analytic as a function of $z_2$ and anti-analytic as a function
of $z_1$.
If $V$ is another domain with smooth boundary and
if $f\colon U\to V$ is a bijective complex analytic map
sending the marked points on $U$ to those of $V$,
and if $F_+^V,F_-^V$ are the functions defined as above for the region $V$
then
\begin{eqnarray}\label{transformation}
F_+^U(v,w)&=&f'(v)F_+^V(f(v),f(w))\\
F_-^U(v,w)&=&\overline{f'(v)}F_-^V(f(v),f(w)).
\end{eqnarray}
\end{prop}
\begin{proof}
We already know that $F_+,F_-$ are analytic in the second variable.
Going back to the coupling function, for a fixed black vertex $v_2$ 
not adjacent to $v_1$ we have 
$$-C(v_1+\eps,v_2)+C(v_1-\eps,v_2)-iC(v_1+i\eps,v_2)+iC(v_1-i\eps,v_2)=0.$$
If $v_2\in B_0$ and $v_1+\eps\in W_0$ this gives in the limit
(using Theorem \ref{longrange})
$$-\partial_{x_1}\Re F_0(v_1,v_2)+\partial_{y_1}\Im F_1(v_1,v_2)=0$$
and if $v_2\in B_1$ and $v_1+\eps\in W_0$ this gives
$$-\partial_{x_1}\Im F_0(v_1,v_2)-\partial_{y_1}\Re F_1(v_1,v_2)=0.$$
These can be combined into a single complex equation
$$-\partial_{x_1}F_0(v_1,v_2)-i\partial_{y_1} F_1(v_1,v_2)=0.$$
Similarly if $v_1+\eps\in B_1$ this gives
$$-\partial_{x_1}F_1(v_1,v_2)-i\partial_{y_1}F_0(v_1,v_2)=0.$$
Summing these gives $\partial_{\overline{z_1}}(F_0+F_1)=0$
and taking their difference and conjugating gives
$\partial_{z_1}(F_0-F_1)=0$. This proves the first two statements.

As a function of $z_2$, the 
function $F_0^V(f(z_1),f(z_2))$ has all the properties of $F_0^U$ except
that the residue at $z_2=z_1$ is $\frac1{\pi f'(z_1)}$.
Similarly the 
function $F_1^V(f(z_1),f(z_2))$ has all the properties of $F_1^U$ except
that the residue at $z_2=z_1$ is $\frac1{\pi f'(z_1)}$.
So letting $\alpha,\beta$ be the real and imaginary parts of $f'(z_1)$ we have that
$$\alpha(z_1)F_0^V(f(z_1),f(z_2))+i\beta(z_1)F_1^V(f(z_1),f(z_2))$$
has residue $\frac{\alpha(z_1)+i\beta(z_1)}{\pi f'(z_1)}=\frac1\pi$
at $z_2=z_1$, and all the other properties of $F_0^U$, and
so must equal $F_0^U$ since $F_0^U$ is unique. A similar argument
shows that 
$$i\beta(z_1)F_0^V(f(z_1),f(z_2))+\alpha(z_1)F_1^V(f(z_1),f(z_2))=F_1^U.$$

The equations for $F_+$ and $F_-$ follow.
\end{proof}

As an example, on the upper half plane we have from (\ref{polescancel})
and (\ref{polesdontcancel}) that
$$F_0(z_1,z_2)=\frac1{\pi(z_2-z_1)}- \frac1{\pi(z_2-\overline{z_1})},$$
and
$$F_1(z_1,z_2)=\frac1{\pi(z_2-z_1)}+ \frac1{\pi(z_2-\overline{z_1})}.$$
These functions vanish at $\infty$, which can be thought of as the 
location of $d_0$.
In particular $F_+(z_1,z_2)=\frac2{\pi(z_2-z_1)}$, which is analytic
in both variables, and $F_-(z_1,z_2)=-\frac2{\pi(z_2-\bar z_1)}$,
which is analytic in $z_2$ and antianalytic in $z_1$.

Let $U$ be the upper half plane with $d_0$ located at $0$ (that is, a square of
type $B_1$ is removed near the origin).
We can compute $F_0^U,F_1^U$ for this new region $U$ 
by using the above transformation rules.
A conformal isomorphism from the upper half plane to itself which takes $0$
to $\infty$ is $f(z)=-1/z$.

Since $f'(z_1)=z_1^{-2}$ we have
$$F^U_+(z_1,z_2)=\frac1{z_1^2}\frac2{\pi(f(z_2)-f(z_1))}$$
$$=\frac{2z_2}{\pi z_1(z_2-z_1)}.$$
Any other choice of $f(z)$ would give the same result.
The function $F^U_-$ is obtained similarly.

\begin{lemma}\label{detneq0} Suppose $\delta>0$.
If $Q$ is an $n\times n$ matrix $Q=(q_{ij})$ and for all $i$,
$$q_{ii}-\delta >\sum_{j,~j\neq i}|q_{ji}|$$
then $\det{Q}>\delta^n>0$.
\end{lemma}
\begin{proof}
Gaussian elimination using rows preserves this property: if for each $j$ we multiply
the first row by $q_{j1}/q_{11}$ and subtract it from the
$j$th row, the first column of the new matrix is all $0$ except for the first
entry $q_{11}$, and the remaining $n-1\times n-1$ submatrix 
still has the property in the statement. For example the first
column of the submatrix is 
$$\left(q_{22}-\frac{q_{12}}{q_{11}}q_{21},~
q_{32}-\frac{q_{12}}{q_{11}}q_{31},\ldots,~
q_{n2}-\frac{q_{12}}{q_{11}}q_{n1}\right),
$$
and 
\begin{eqnarray*}
q_{22}-\frac{q_{12}}{q_{11}}q_{21}-\delta&>&
|q_{12}|+|q_{32}|+\ldots+|q_{n2}|-\frac{q_{12}}{q_{11}}q_{21}\\
&\geq&|q_{32}|+\dots+|q_{n2}|+\left(|q_{12}|-
\frac{|q_{12}|\cdot|q_{21}|}{q_{11}}\right)\\
&=&|q_{32}|+\dots+|q_{n2}|+|q_{12}|\left(\frac{
q_{11}-|q_{21}|}{q_{11}}\right)\\
&>&|q_{32}|+\dots+|q_{n2}|+\frac{|q_{12}|}{q_{11}}\left(
\delta+|q_{31}|+\dots+|q_{n1}|\right)\\
&\geq&|q_{32}-\frac{q_{12}}{q_{11}}q_{31}|+\dots+
|q_{n2}-\frac{q_{12}}{q_{11}}q_{n1}|.
\end{eqnarray*}
\end{proof}

Recall that the continuous Green's function on a region $U$
is the real-valued function $g_U$ 
satisfying $\Delta g_U(z_1,z_2)=\delta_{z_1}(z_2)$, and
which is zero when $z_2$ is on the domain boundary
(here $\delta_{z_1}$ is the {\it continuous} delta-function,
and $\Delta=-\frac{\partial^2}{\partial x^2}-\frac{\partial^2}{\partial y^2}$).

\begin{lemma} \label{GGconv}
Let $z_1=x_1+iy_1$ be a point in the interior of $U$, and let $z_2\in U$, $z_2\neq z_1$. 
Let $v_1$ be a vertex
of $\Bo'(P_\eps)$ within $O(\eps)$ of $z_1$, and let $v_2$ be a vertex
of $\Bo'(P_\eps)$ within $O(\eps)$ of $z_2$.
Then the difference of 
(rescaled) Green's functions $\frac1{\epsilon}G(v_1+\eps,v_2)
-\frac1\eps G(v_1-\eps,v_2)$ converges
to $2\partial_{x_1} g_U(z_1,z_2)$.
\end{lemma}

\begin{proof}
Let $H(v_1,v_2)=\frac1\eps(G(v_1+\eps,v_2)-G(v_1-\eps,v_2))$.
From Theorem \ref{1/z}, on the plane $\eps\Z^2$ we have
$$H_0(v_1,v_2)\stackrel{{\rm def}}{=}\frac1\eps(G_0(v_1+\eps,v_2)-
G_0(v_1-\eps,v_2))=
\Re\frac1{\pi (v_2-v_1)}+O(\frac1{|v_2-v_1|^2}).$$

The function $H(v_1,v_2)-H_0(v_1,v_2)$ is harmonic (as a function of $v_2$)
on all of $\Bo'(P_\eps)$ (including $v_1\pm\eps$) and
has bounded boundary values, since $H_0(v_1,v_2)$ is $O(1)$
on the boundary of $\Bo'(P_\eps)$ and $H(v_1,v_2)$ is zero there.
Let $g$ be the continuous 
harmonic function which has boundary values equal to the boundary values
of the limit $$\lim_{\eps\to 0}H(v_1,v_2)-H_0(v_1,v_2).$$
Since these boundary values are continuous in the limit, $g$ exists
and is unique. 
Note that the boundary values of $H-H_0$ are within
$O(\eps)$ of the limiting values (Theorem \ref{1/z}).

Restrict $g$ to a 
function on the vertices of $\Bo'(P_\eps)$. The discrete Laplacian of $g$
at a vertex $v\in \Bo'(P_\eps)$ is:
$$\Delta_\eps g(v_1,v)=
4g(v)-g(v+\epsilon)-g(v-\epsilon)-g(v-i\epsilon)-g(v+i\epsilon)$$
and when $\epsilon$ is small 
we can approximate this using the Taylor expansion of the smooth
function $g$, yielding
$$\Delta_\eps g(v_1,v)=-\frac{\epsilon^4}{24}
\left(\frac{\p^4g(v)}{\p x^4}+\frac{\p^4g(v)}{\p y^4}
\right)+O(\epsilon^5).$$

Therefore $H(v_1,v_2)-H_0(v_1,v_2)-g(v_1,v_2)$ 
has discrete Laplacian which is $O(\epsilon^4)$ on $\Bo'(P_\eps)$,
and the boundary values are $O(\eps)$.
A standard argument now shows that $H-H_0$ is close to $g$:
the function $x+iy\mapsto x^2$ 
has discrete Laplacian which is a constant;
choose constants $B_2,B_3$ sufficiently large so that
$$\Delta_\eps\left(B_2\epsilon^4(\Re(v_2))^2+H(v_1,v_2)-H_0(v_1,v_2)-
g(v_1,v_2)\right)\geq 0$$
and
$$\Delta_\eps\left(B_3\epsilon^4(\Re(v_2))^2-H(v_1,v_2)+H_0(v_1,v_2)+
g(v_1,v_2)\right)\geq 0$$
on $\Bo'(P_\eps)$.
By the maximum principle for superharmonic functions, 
these functions must take their maximum value
on the boundary of the domain $\Bo'(P_\eps)$. 
Since $H(v_1,v_2)-H_0(v_1,v_2)-g(v_1,v_2)=O(\eps)$ 
on the boundary of $\Bo'(P_\eps)$, we conclude that
$$|H(v_1,v_2)-H_0(v_1,v_2)-g(v_1,v_2)|=O(\epsilon).$$
Therefore $H(v_1,v_2)$ converges to the function 
$\Re\frac1{\pi(v_2-v_1)}+g(v_1,v_2)$ which has boundary values $0$ and a single
``pole'' of residue $1/\pi$ at $v_1$. This
is $2$ times the $x_1$-derivative of the continuous Green's function.
\end{proof}

A similar result holds for the $y_1$-derivative of $g_U$, yielding:
\begin{cor}\label{dx} Recall the definitions of the functions $F_0,F_1,F_+,F_-$ from Theorem \ref{longrange}
and Proposition \ref{Finvariant}.  Letting $z_1=x_1+iy_1$, we have 
$$2dg_U(z_1,z_2) = F_0(z_1,z_2)dx_1 + F_1(z_1,z_2)dy_1 = \frac12 F_+(z_1,z_2)dz_1 + \frac12 F_-(z_1,z_2)\overline{dz_1}$$
where the exterior differentiation $dg_U$ is with respect to the first variable.
\end{cor}

When $z_1\in\partial U$ the proof of Lemma \ref{GGconv} implies the convergence of
the Green's function as well.
\begin{cor}\label{bdygreen}Let $\delta>0$.
If $z_1$ is on the boundary of $U$, and the boundary of both $U$ and $P_\eps$
is straight and horizontal in a $\delta$-neighborhood of $z_1$, 
then for $v_1$ within $O(\eps)$ of $z_1$, 
$$\frac1\eps G(v_1,v_2)=g_U(z_1,z_2)+o(1).$$
\end{cor}
\begin{proof}
Reflect $\Bo'(P_\eps)$
across the boundary edge near $z_1$ (the edge consisting of vertices in $Y$)
to get a graph $\Bo''(P_\eps)$. Glue $\Bo'(P_\eps)$
and $\Bo''(P_\eps)$ along their common edge in a $\delta$-neighborhood of $v_1$.
A harmonic function $f$ on $\Bo'(P_\eps)$ which is zero on the boundary
extends to a harmonic function on this glued graph by setting
$f(v')=-f(v)$ when $v'$ is the reflection of $v$.
In other words the Green's function $G(v_1,v_2)$ on $\Bo'(P_\eps)$ is 
the difference of two Green's functions on $\Bo'(P_\eps)\cup \Bo''(P_\eps)$;
one centered at $v_1$ and one centered at $v_1'$.

On the glued graph $\Bo'(P_\eps)\cup \Bo''(P_\eps)$, the vertices $v_1,v_1'$ 
are at distance at least $\delta$ from the boundary $\partial(\Bo'(P_\eps)\cup \Bo''(P_\eps))$, but only distance
$O(\eps)$ from each other.
The argument of Lemma \ref{GGconv} can then be applied in this case,
replacing $H(v_1,v_2)$ by $\frac1\eps(G(v_1,v_2)-G(\overline{v_1},v_2))$.
\end{proof}

A similar result holds when the boundary is vertical.

\section{Conformal invariance of heights}\label{1proof}
\subsection{Proof of Theorem \protect{\ref{1}}} 
Let $U$ be a region in $\C$ with boundary which is piecewise smooth as previously defined.
Let $d_j'$ be a point on the $j$-th boundary component $D_j$ of $U$.
Let $e_j'\neq d_j'$ be another point of $D_j$, which is not at a corner of the boundary.

Let $P_\eps$ be a Temperleyan polyomino 
approximating $U$ in the sense of section \ref{approx}, with the additional
constraint of having horizontal boundary in a neighborhood of each $e_j'$,
and so that the interior of $U$ is locally below each $e_j$.
We show that the distribution of the heights of the
boundary components of $P_\eps$ is conformally invariant.

Let $e_j$ be a vertex on the boundary of $P_\eps$ near $e_j'$.
We assume for simplicity that each $e_j$ has the same parity (its coordinates have the same parity)
as $e_0$.  For definiteness we suppose the lattice square whose lower
left corner is $e_j$ is of type $B_1$ for each $j$.

Let $h_j$ be the random variable giving the height of $e_j$ for a random
tiling of $P_\eps$ assuming the height of $e_0$ is zero.
Let $\bar h_j$ be the mean value of $h_j$.

We will show that for integers $n_1,n_2,\ldots,n_k\geq 0$, the moment
\begin{equation}
\E((h_1-\bar h_1)^{n_1}(h_2-\bar h_2)^{n_2}\cdots
(h_k-\bar h_k)^{n_k})\label{moment2}
\end{equation}
is conformally invariant.
Let $K=n_1+\dots+n_k$.
The precise value of the moment (\ref{moment2}) is as follows.
\begin{prop}\label{theformula}
Let $\{\gamma_i\}_{i\in[1,K]}$ be a collection of pairwise disjoint paths in $U$, 
such that for each $j\in[1,k]$ there are 
$n_j$ paths runnning from the outer boundary to the $j$th boundary component.
Then as $\eps\to0$ the moment (\ref{moment2}) converges to 
\begin{equation}\label{mom}
\sum_{\vep_1,\dots,\vep_K\in\{\pm1\}}\vep_1\cdots \vep_K\int_{\gamma_1}\cdots\int_{\gamma_{K}}\det_{i,j\in[1,K]}\Bigl(
F_{\vep_i,\vep_j}(z_i,z_j)\Bigr)dz_1^{(\vep_1)}\cdots 
dz_K^{(\vep_k)},
\end{equation}
where $dz_j^{(1)}=dz_j$ and $dz_j^{(-1)}=d\overline{z_j}$, and 
$$F_{\vep_i,\vep_j}(z_i,z_j)=\left\{\begin{array}{ll}
0&\mbox{\rm if }i=j\\
F_+(z_i,z_j)&\mbox{\rm if }(\vep_i,\vep_j)=(1,1)\\
F_-{(z_i,z_j)}&\mbox{\rm if }(\vep_i,\vep_j)=(-1,1)\\
\overline{F_-{(z_i,z_j)}}&\mbox{\rm if }(\vep_i,\vep_j)=(1,-1)\\
\overline{F_+{(z_i,z_j)}}&\mbox{\rm if }(\vep_i,\vep_j)=(-1,-1).\end{array}\right.$$
\end{prop}

Note that in each of the $2^K$ multiple integrals in (\ref{mom}), the integrand $I$ is conformally invariant,
in the sense that 
$$\int_\gamma I({\bf z})d{\bf z} = \int_{f(\gamma)}I(f(\bf z))d\bf z.$$
This follows because of the transformation rules (\ref{transformation}) and the fact that each
integrand is analytic or antianalytic in $z_i$ according to $\vep_i=\pm 1$.
Therefore the moment (\ref{moment2}) is conformally invariant.

An example calculation is done in section \ref{2mom}.

By \cite[section 30]{Bil}, there is a unique probability distribution with these moments
on condition that the moment generating function
$$H(t_1,\dots,t_k)=\sum_{n_1,\dots,n_k\geq 0} \frac{m(n_1,\dots,n_k)t_1^{n_1}\cdots t_k^{n_k}}{n_1!\cdots n_k!}$$
has nonzero radius of convergence around the origin (here $m(n_1,\dots,n_k)$ is a shorthand
for (\ref{moment2})). This convergence is shown in Lemma \ref{momgenconv}, below.
We can then conclude that the probability distribution with these moments is conformally invariant, and
by \cite[Theorem 30.2]{Bil} that this distribution is the limit of the distributions for finite $\eps$.
This will complete the proof of Theorem \ref{1}.

\noindent{\bf Proof of Proposition \ref{theformula}.}

For each $\eps$ sufficiently small and 
for each $j\in[1,k]$ let $\gamma_{j1}^{(\eps)},\ldots,\gamma_{jn_j}^{(\eps)}$ 
be pairwise disjoint lattice paths (which are also disjoint for distinct $j$s)
in $P_\eps$ which start on the flat boundary near $e_0$ and end on the flat boundary near $e_j$ .
We require that each straight edge of 
$\gamma_{js}^{(\eps)}$ have even length (by this we mean, a 
length which is an even multiple of $\eps$).
This is possible by our choice of parities for $e_0$ and $e_j$.

In a given tiling the height change on $\gamma^{(\eps)}_{js}$ equals
$4(A_{js}-B_{js})$, where $A_{js}$ 
is the number of dominos crossing $\gamma^{(\eps)}_{js}$ with the black
square on the right and $B_{js}$ is the number of dominos crossing 
$\gamma_{js}^{(\eps)}$
with the black square on the left. To see this, note that if 
$\gamma^{(\eps)}_{js}$ does not cross any dominos,
the height change is $0$: the straight edges have even length so the
height change along them is zero. Then, for each domino crossed by 
$\gamma^{(\eps)}_{js}$, the height difference changes along that edge from $-1$ to $+3$
if the domino has
black square on the right, and from $+1$ to $-3$ if the black square is on the left.

Since $h_j=4(A_{js}-B_{js})$ for each $s$,
the moment (\ref{moment2}) is equal to 
\begin{equation}\label{AB}
4^K\E\Bigl((A_{11}-B_{11}-\bar A_{11}+\bar B_{11})\cdots(A_{kn_k}-B_{kn_k}-\bar A_{kn_k}
+\bar B_{kn_k})\Bigr)
\end{equation}
where $K=n_1+\dots+n_k$.

The remainder of the proof involves expanding this out, cancelling various
terms and then recombining in the right way. 

For notational simplicity we renumber the paths $\gamma^{(\eps)}_{js}$
from $1$ to $K$. Similarly change indices of $A_{js},B_{js}$
to values in $[1,K]$. For $j\in[1,K]$ let $\alpha_{jt}$ 
be the $t$-th possible domino of $\gamma^{(\eps)}_{j}$
crossing $\gamma^{(\eps)}_{j}$ whose black
square is right of $\gamma^{(\eps)}_{j}$. Similarly let $\beta_{jt}$ be the
$t$-th possible domino crossing $\gamma^{(\eps)}_{j}$ 
whose black square is on the left.
Let $\alpha_{jt},\beta_{jt}$ also denote the indicator functions of the presence of these
edges/dominos.
Then 
\begin{equation}\label{A-B}
A_{j}-B_{j}=\sum_t \alpha_{jt}-\sum_{t'} \beta_{jt'}.
\end{equation}

Let $(w_{js},b_{js})$ be the white and black squares,
respectively, of the domino $\alpha_{js}$ and 
$(w_{js}',b_{js}')$ be the white and black squares of the domino $\beta_{js}$.

Since the straight edges in the path $\gamma^{(\eps)}_{j}$ have even length,
we can pair the $\alpha_{jt}$ dominos with adjacent $\beta_{jt'}$ dominos which are parallel
to $\alpha_{jt}$.
It is then convenient to write 
$$A_j-B_j-\bar A_j+\bar B_j=\sum_t (\alpha_{jt}-\bar \alpha_{jt}-\beta_{jt}+
\bar \beta_{jt})$$ where $\alpha_{jt}$ and $\beta_{jt}$ are paired.
Equation
(\ref{AB}) is now
\begin{equation}\label{sum}
4^K\sum_{t_1,\dots,t_\ell}\E\Bigl((\alpha_{1t_1}-\bar \alpha_{1t_1}-\beta_{1t_1}+\bar \beta_{1t_1})
\cdots (\alpha_{Kt_K}-\bar \alpha_{K t_K}-\beta_{K t_K}+\bar \beta_{K t_K})
\Bigr),\end{equation}
where the sums are over all pairs $(\alpha_{1t_1},\beta_{1t_1})$ of $\gamma^{(\eps)}_{1}$, 
$(\alpha_{2t_2},\beta_{2t_2})$ of $\gamma^{(\eps)}_{2}$
and so on.

\begin{lemma}\label{0diag}
Let $e_i=(w_i,b_i)$ for $i=1,\dots,n$ be a set of $n$ disjoint edges; then 
$$\E((e_1-\bar e_1)\cdots(e_n-\bar e_n))=a_E\det\left(\begin{array}{cccc}
0&C(w_1,b_2)&\dots&C(w_1,b_n)\\C(w_2,b_1)&0&&\vdots\\\vdots&&&C(w_{n-1},b_n)\\C(w_n,b_1)&\dots&
C(w_n,b_{n-1})&0\end{array}
\right),$$
where (using the convention after Theorem \ref{ken}) $a_E$ is the product of the edge weights of the $e_i$. 
\end{lemma}
\begin{proof}
This follows from Theorem \ref{ken}, induction on $n$ and the fact that
$$\left|\begin{array}{cccc}0&a_{12}&\dots&a_{1n}\\a_{21}&a_{22}&&\\\vdots&&\ddots&\\a_{n1}&&&a_{nn}\end{array}\right|=
\left|\begin{array}{cccc}a_{11}&a_{12}&\dots&a_{1n}\\a_{21}&a_{22}&\\\vdots&&\ddots&\\a_{n1}&&&a_{nn}\end{array}\right|-
\left|\begin{array}{cccc}a_{11}&0&\dots&0\\0&a_{22}&\dots&a_{2n}\\\vdots&\vdots&&\vdots\\0&a_{n2}&\dots&a_{nn}\end{array}\right|.$$
\end{proof}
Now expand the summand of (\ref{sum}) into $2^K$ terms
\begin{equation}\label{2^K}
\E((\alpha_{1t_1}-\bar\alpha_{1t_1})\dots(\alpha_{K t_K}-\bar\alpha_{K t_K}))+\ldots+(-1)^K
\E((\beta_{1t_1}-\bar\beta_{1t_1}) \dots(\beta_{K t_K}-\bar\beta_{K t_K})).
\end{equation}
By Lemma \ref{0diag}, each
term is a certain quantity $a_E$ times the determinant of a $K\times K$
matrix whose entries are given by the coupling function
connecting black squares of the dominos $\alpha_{st_s},\beta_{st_s}$ 
with white squares of the other dominos.
Since each `$\beta$' edge has weight of the opposite sign as the `$\alpha$'
edge to which it is paired, 
the signs in (\ref{2^K}) cancel with the sign changes in the $a_E$ and so
(\ref{2^K}) is equal to the {\it sum}
of all $2^K$ determinants, times the product $a_E$ of the edge weights of the {\it first} determinant.

Consider the first term in (\ref{2^K})
\begin{equation}\label{Eas}\E((\alpha_{1t_1}-\bar\alpha_{1t_1})\dots(\alpha_{K t_K}-\bar\alpha_{K t_K})).
\end{equation}

Recall that $(w_{js},b_{js})=\alpha_{js}$ and $(w_{js}',b_{js}')=\beta_{js}$.
Fix a choice of indices $s=s_j$ for the moment so we can drop the second subscripts.
By Lemma \ref{0diag}, equation (\ref{Eas}) is then equal to 
\begin{equation}\label{deta}
a_E\left|\begin{array}{cccc}0&C(w_2,b_1)&\dots&C(w_K,b_1)\\
C(w_1,b_2)&0&&\vdots\\\vdots &&\ddots&C(w_K,b_{K-1})\\ C(w_1,b_K)&\ldots&C(w_{K-1},b_K)&0\end{array}\right|.
\end{equation}
A typical term in the expansion of (\ref{deta}) is 
\begin{equation}\label{sigma}
a_E\mbox{sgn}(\sigma)C(w_1,b_{\sigma(1)})C(w_2,b_{\sigma(2)})\cdots C(w_K,b_{\sigma(K)})
\end{equation}
where $\sigma$ has no fixed points.

Let us first assume that $\sigma$ is a $K$-cycle; reorder the indices so that (\ref{sigma})
becomes
\begin{equation}\label{cycle}
a_E\mbox{sgn}(\sigma)C(w_1,b_2)C(w_2,b_3)\cdots C(w_K,b_1).
\end{equation}

To expand this out, define variables $r_i=\pm 1$ according to whether $w_i\in W_0$ or $w_i\in W_1$,
and $s_i=\pm1$ according to whether $b_i\in B_0$ or $b_i\in B_1$.
If we assume that neither $w_1$ or $b_2$ is close to the boundary, 
we can then write (see Theorem \ref{longrange} and the remarks immediately after its statement)
\begin{eqnarray*}
C(w_1,b_2) &=& \eps\left(\frac{1-r_1s_2}2i\Im + \frac{1+r_1s_2}2\Re\right)
\left(\frac{1+r_1}2F_0(w_1,b_2)+\frac{1-r_1}2F_1(w_1,b_2)\right)+o(\eps)\\
&=&\frac{\eps}4(F_+(w_1,b_2)+r_1F_-(w_1,b_2)+s_2\overline{F_-}(w_1,b_2)+r_1s_2\overline{F_+}(w_1,b_2))+o(\eps).
\end{eqnarray*}

For each fixed $\xi>0$, when neither of $w_1,b_2$ are within $\xi$ of the boundary, this approximation
holds for sufficiently small $\eps$. When one or both of $w_1,b_2$ are within $\xi$ of the boundary, we only need to know
that $\frac1\eps C(w_1,b_2)$ is bounded by some constant independent of $\eps$ and $\xi$.
Then in the sum (\ref{sum}) (and in the integral (\ref{mom}) we can ignore all terms
in which some $w_i$ or $b_j$ is within $\xi$ of the boundary, as these will
contribute at most $O(\xi)$. The boundedness of $\frac1\eps C(w_1,b_2)$ follows
from the convergence of the discrete Green's function as in Theorem \ref{longrange}.

We can now write (\ref{cycle}) as
\begin{eqnarray}\label{4cyc}
4^{-K}\eps^Ka_E\mbox{sgn}(\sigma)
\Bigl((F_+(w_1,b_2)+r_1F_-(w_1,b_2)+s_2\overline{F_-}(w_1,b_2)+r_1s_2\overline{F_+}(w_1,b_2))\cdots\hskip1in\\
\nonumber
\hskip1in(F_+(w_K,b_1)+r_KF_-(w_K,b_1)+s_1\overline{F_-}(w_K,b_1)+r_Ks_1\overline{F_+}(w_K,b_1))
\Bigr)+o(\eps^K).
\end{eqnarray}
We obtain a similar expression if we replace $(w_1,b_1)$ by $(w_1',b_1')$, except that
the signs of $r_1$ and $s_1$ are reversed. In particular if we sum up over all $2^K$ choices
of $\alpha_j$ and $\beta_j$ (as we need to do to obtain (\ref{2^K})), we get $2^K$ times 
the sum of those terms in (\ref{4cyc}) which have $r_i$ to the same power ($1$ or $0$) as $s_i$, for each $i$.
This sum can therefore be written as an error $o(\eps^K)$ plus
\begin{equation}\label{Fsum}
2^{-K}\eps^K\mbox{sgn}(\sigma)a_E
\sum_{\vep_1,\dots,\vep_K\in\{-1,1\}}(r_1s_1)^{(1-\vep_1)/2}\cdots(r_Ks_K)^{(1-\vep_K)/2}F_{\vep_1,\vep_2}(z_1,z_2)F_{\vep_2,\vep_3}
(z_2,z_3)\cdots F_{\vep_K,\vep_1}(z_K,z_1),
\end{equation}
where 
$F_{\vep_i,\vep_j}(z_i,z_j)$ is as defined in Proposition \ref{theformula}. 

Now in view of replacing the sum (\ref{2^K}) by an integral when $\eps$ is small, we
can replace $\eps$ by a certain phase times $\frac12 dz_j$ or $\frac12 d\bar z_j$. 
When the path $\gamma_j$ is going east (horizontal and to the right), 
we have $2\eps = dx_j = dz_j = d\bar z_j,$ and the edge of type $\alpha$ has weight
$-i$, because its upper vertex is white and lower vertex black (recall that edges of type $\alpha$
have black vertices on their right). Furthermore $r_js_j=-1$ on an east-going path.
When the path $\gamma_j$ is going west, $2\eps = -dx_j = -dz_j = -d\bar z_j$, the edge of type $\alpha$
has weight $i$, and $r_js_j=-1$.
When the path $\gamma_j$ is going north, $2\eps = dy_j = -idz_j = id\bar z_j$, the edge $\alpha$
has weight $1$, and $r_js_j=1$.
When the path $\gamma_j$ is going south, $2\eps = -dy_j = idz_j = -id\bar z_j$, the edge $\alpha$
has weight $-1$, and $r_js_j=1$. 
Notice that in each case $2\eps$ times the edge weight, times $(r_js_j)^{(1-\vep_j)/2}$ is
$-\vep_jidz_j^{(\vep_j)}$
(recall the definition of $dz_i^{(\vep_i)}$ from Proposition \ref{theformula}).
Recalling that $a_E$ is the product of the 
edge weights (of the $\alpha$-type edges), for any choices of the $\vep_j$ we have 
$$a_E(2\eps)^K(r_1s_1)^{(1-\vep_1)/2}\cdots(r_Ks_K)^{(1-\vep_K)/2} = (-i)^K\vep_1\cdots\vep_Kdz_1^{(\vep_1)}\cdots dz_K^{(\vep_K)}.$$
The sum (\ref{Fsum}) is therefore
\begin{equation}\label{lastsum}
4^{-K}(-i)^K\mbox{sgn}(\sigma)
\sum_{\vep_1,\dots,\vep_K\in\{-1,1\}}\vep_1\cdots\vep_K F_{\vep_1,\vep_2}(z_1,z_2)F_{\vep_2,\vep_3}
(z_2,z_3)\cdots F_{\vep_K,\vep_1}(z_K,z_1)dz_1^{(\vep_1)}\cdots dz_K^{(\vep_K)}.
\end{equation}

When $\sigma$ is a product of disjoint cycles we can treat each cycle
separately and the result is the product of terms like (\ref{lastsum}) involving
disjoint sets of indices. Thus when we sum over all (fixed-point free) permutations we obtain the formula
of the proposition, but without the integral. The factor of $4^{-K}$ cancels with
the factor of $4^K$ in (\ref{sum}), and summing over all pairs gives the integral
in (\ref{mom}). This completes the proof.

\begin{lemma}\label{momgenconv} The moment generating function for the moments (\ref{mom}) has
positive radius of convergence.
\end{lemma}

\begin{proof}
Letting $K=n_1+\dots+n_k$ denote the ``size'' of the moment, 
it suffices to show that a moment of size $K$ is smaller than $(cK)^K$ for a constant $c$.
Let $\gamma_1,\dots,\gamma_K$ be the paths of integration in (\ref{mom}).
We can choose the $\gamma_i$ so that no two are closer than  $c_1/K$ for some constant $c_1$;
indeed, we can choose the paths so that the distance between $\gamma_i$ and $\gamma_j$ is at
least $c_1|i-j|/K$. Since $F_0(z_1,z_2)$ and $F_1(z_1,z_2)$ are $O(\frac1{|z_1-z_2|})$,
in the determinant in (\ref{mom}) the $ij$-entry is at most
$c_2 K/|i-j|$ in absolute value. The determinant of a matrix is bounded by the product
of the $\ell_2$-norms of its rows, and each row of the determinant in (\ref{mom}) has $\ell_2$-norm
bounded by $K(2+\frac{2}{2^2}+\dots+\frac{2}{(K/2)^2})^{1/2}=c_3 K$ for another constant $c_3$. 
Therefore the sum of the integrals in (\ref{mom}) is bounded by $c_4^K K^K$ for a constant
$c_4$.  This completes the proof.
\end{proof}

\subsection{The average height.}\label{avght}
Let $U\subset\C$ be a region with piecewise smooth boundary as previously defined.
Let $b$ be a point on the outer boundary of $U$, $b\neq d_0$.
For each $\eps\ll\delta$ let $P_\eps$ approximate $U$ as in section \ref{approx},
but with the additional constraint of having horizontal 
boundary in a $\delta$-neighborhood of $b$.
(We also assume that the interior of $P_\eps$ is locally above the boundary at $b$.)
Let $z$ be a point in the interior of $U$. 
Let $z'\in P_\eps$ be within $O(\eps)$ of $z$
and let $b'\in\partial P_\eps$ be within $O(\eps)$ of $b$.
We assume that $b'$ and $z'$ are the lower left corners of lattice squares of type $B_1$.
Let $\gamma^{(\eps)}$ be a lattice path from $b'$ to $z'$
such that all edges of $\gamma^{(\eps)}$ have even length, and which starts straight and
northgoing for a distance at least $c\delta$ for some constant $c$.
In the notation of the previous section, we have
$\E(h(z))=4\sum\E(\alpha_s)-\E(\beta_s)$ where $\alpha_s,\beta_s$ are pairs of potential
dominos crossing the path $\gamma^{(\eps)}$.

Near the boundary, $\gamma^{(\eps)}$ is northgoing. When $\alpha_s,\beta_s$ are within
$o(1)$ of the boundary we have
$$\E(\alpha_s)=C(w_s,b_s)=C_H(w_s,b_s)+O(\eps)$$
and $$\E(\beta_s)=-C(v'_s,w'_s)=-C_H(w_s',b_s')+O(\eps)$$
(note that $\alpha_s$ has weight $1$ and $\beta_s$ has weight $-1$ when $\gamma^{(\eps)}$ is northgoing).
Therefore using Theorem \ref{nearbdy}
\begin{eqnarray*}
\E(\alpha_s-\beta_s)&=&C_H(w_s,b_s)+C_H(w_s',b_s')+O(\eps)\\
&=&\frac14+C_0(\overline{w_s},b_s)+\frac{-1}4+
C_0(\overline{w_s'},b_s')+O(\eps)\\
&=&C_0(0,b_s-\overline{w_s})+C_0(0,b_s'-\overline{w_s'})+O(\eps).
\end{eqnarray*}
Near the boundary, $b_s-\overline{w_s}$ takes successively values $1+2i,1+6i,\dots,1+(2+4k)i\dots$
and $b_s'-\overline{w_s'}$ takes successively values $1+4i,1+8i,\dots,1+4ki\dots$.
When we sum over all pairs $(w_s,b_s),(w_s',b_s')$ on the path
$\gamma^{(\eps)}$ which are within $o(1)$ of the boundary, the contribution is $o(\eps)$ plus 
$$C_0(0,1+2i)+C_0(0,1+4i)+\dots+C_0(0,1+2ki)+\dots=\frac12.$$
(This formula can be proved analytically from Proposition (\ref{C0form})
or more simply by symmetry, noting that the average height on the upper half-plane
is $\frac12$ given that the height on the boundary alternates between $0$ and $1$.)

For the terms not near the boundary
we have, by Theorem \ref{longrange}, when $\gamma^{(\eps)}$ is northgoing, 
\begin{eqnarray*}
C(w_s,b_s)+C(w_s',b_s')&=&\frac14+\eps\Re F_1^*(z_s,z_s) + \frac{-1}4+\eps
\Re F_0^*(z_s,z_s)+o(\eps)\\
&=&\Re(F_+^*(z_s,z_s)\eps)+o(\eps)
\end{eqnarray*}
where $z_s$ is the coordinate of $w_s$ and $F_+^*=F_0^*+F_1^*$. 
Similarly for the other directions of $\gamma$ we have 
$$C(w_s,b_s)+C(w_s',b_s')=
\left\{\begin{array}{ll}
\Re(-F_+^*(z_s,z_s)\eps)+o(\eps)&\mbox{when }\gamma
\mbox{ is southgoing}\\
\Im(F_+^*(z_s,z_s)\eps)+o(\eps)&\mbox{when }\gamma 
\mbox{ is eastgoing}\\ 
\Im(-F_+^*(z_s,z_s)\eps)+o(\eps)&\mbox{when }\gamma 
\mbox{ is westgoing}.\end{array}\right.$$

We can replace $\eps$ by $\frac12dz_s,-\frac{i}2dz_s,
-\frac12dz_s,\frac{i}2dz_s$ respectively
according to whether $\gamma$ is east-, north-, west-, or southgoing.
Then all four cases become
$$C(w_s,b_s)+C(w_s',b_s')=\frac12\Im\left(F_+^*(z_s,z_s)dz_s\right)+o(\eps).$$

The average height is then given by the imaginary part of
the integral of $2F_+^*(z,z)dz$ from $b$ to $z$
(recall the factor of $4$ from the first paragraph of this section), plus $\frac12$, the constant coming from the boundary.
This expression does not depend on $\delta$.

For another region $V$ conformally equivalent to $U$
we have the following. Let $f\colon V\to U$ be a conformal
isomorphism. Then $F_+^V(z_1,z_2)=f'(z_1)F_+^U(f(z_1),f(z_2))$ from Proposition \ref{Finvariant}.
Therefore
\begin{eqnarray*}
(F_+^V)^*(z_1,z_2)&=&F_+^V(z_1,z_2)-\frac2{\pi(z_2-z_1)}\\
&=&-\frac2{\pi(z_2-z_1)}+f'(z_1)\left((F_+^U)^*(f(z_1),f(z_2))+
\frac2{\pi(f(z_2)-f(z_1))}\right)
\end{eqnarray*}
and in the limit as $z_2\to z_1$ this is (simplifying 
using the Taylor expansion of $f$)
$$f'(z_1)(F_+^U)^*(f(z_1),f(z_1))-\frac{f''(z_1)}{\pi f'(z_1)}.$$
So the average height of $z\in V$ equals the average height
of $f(z)$ in $U$, plus a term
$$-\frac2{\pi}\int_{f(\gamma)} (\log f'(z))'dz.$$
This term is $-\frac2\pi$ times the change in total turning (in radians) of
the path $f(\gamma)$ from the path $\gamma$. 

This implies that
if the path $\gamma$ starts at the outer boundary of $U$, at a
point where the tangent vector (chosen in the counterclockwise direction)
has angle $\theta$ with respect to the horizontal axis (where $\theta\in[0,2\pi)$),
then the average height of a point $z\in U$ is 
$$\frac12+\frac{2\theta}{\pi}+2\Im\int_\gamma F_+^*(z_1,z_1)dz_1.$$

Therefore we have
\begin{thm}\label{meanht}
Up to an additive constant, the average height of a point $z$ not within $o(1)$
of the boundary of $U$ is given by
the harmonic function whose boundary values are 
$\frac{2\theta(x)}{\pi}$, where $\theta(x)$ is the total turning (in radians)
of the tangent vector to the boundary on the boundary path going
counterclockwise from $d_0'$ to $x$.
\end{thm}
Note that the boundary values are discontinuous at the point $d_0'$. 

For example,  as noted earlier
on the upper half plane when $d_0=\infty$ the average
height of every point is $\frac12$. When $d_0=0$, rather,
then recall that $F_+(z_1,z_2)=\frac{2z_2}{\pi z_1(z_2-z_1)}$.
So $F_+^*(z_1,z_1)=\frac2{\pi z_1}.$
The average height at a point $z$ is (integrating from $x=\Re(z)$)
\begin{eqnarray*}
\E(h(z))&=&\frac12+2\Im\int_{x}^z\frac{2dz_1}{\pi z_1}\\
&=&\frac12+\frac4{\pi}\Im\log(z/x)\\
&=&\frac12+\frac4{\pi}\arg(z).
\end{eqnarray*}
This is the harmonic function with boundary values (on the axis)
$\frac12$ to the right of the origin and $\frac92=\frac12+4$ to the left
of the origin. Note that on the boundary of the polyomino $P_\eps$,
the height alternates between $0$ and $1$ to the right of the origin and
between $4$ and $5$ to the left of the origin.

\subsection{Example: a second moment computation.}\label{2mom}
For a random tiling of the upper half plane with $d_0=\infty$
we compute the moment $\E((h(p)-\bar h(p))(h(q)-\bar h(q))$ for two points $p,q$.
Since $\bar h(p)=\bar h(q)=\frac12$, this will also give $\E(h(p)h(q))$.

Let $r,s$ be the vertical projections of $p,q$, respectively, to the $x$-axis.
Let $\gamma_1$ and $\gamma_2$ be disjoint paths
running straight from the boundary to $p,q$, respectively.
From Theorem \ref{mom}, we have
$$\E((h(p)-\bar h(p))(h(q)-\bar h(q))=\hskip3in$$
$$\hskip1in=\int_{\gamma_1,\gamma_2}
\left|\begin{array}{cc}0&F_+(z_1,z_2)\\F_+(z_2,z_1)&0\end{array}\right|dz_1dz_2 -\int_{\gamma_1,\gamma_2}
\left|\begin{array}{cc}0&F_-(z_1,z_2)\\\overline{F_-(z_2,z_1)}&0\end{array}\right|d\overline{z_1}dz_2 -\hskip1in$$
$$\hskip1in\int_{\gamma_1,\gamma_2}
\left|\begin{array}{cc}0&\overline{F_-(z_1,z_2)}\\F_-(z_2,z_1)&0\end{array}\right|dz_1d\overline{z_2} +
\int_{\gamma_1,\gamma_2}
\left|\begin{array}{cc}0&\overline{F_+(z_1,z_2)}\\\overline{F_+(z_2,z_1)}&0\end{array}\right|d\overline{z_1}d\overline{z_2}.$$

For the upper half-plane we have $F_+(z_1,z_2)=\frac2{\pi(z_2-z_1)}$ and $F_-(z_1,z_2)=\frac2{\pi(z_2-\overline{z_1})}.$
Plugging these in gives
$$-\frac4{\pi^2}\int_{\gamma_1}\int_{\gamma_2}\frac1{(z_2-z_1)^2}dz_1dz_2 +
\frac4{\pi^2}\int_{\gamma_1}\int_{\gamma_2}\frac1{(z_2-\overline{z_1})^2}
d\overline{z_1}dz_2+
\qquad\qquad$$
$$\qquad\qquad + \frac4{\pi^2}\int_{\gamma_1}\int_{\gamma_2}\frac1{(\overline{z_2}-z_1)^2}
dz_1d\overline{z_2}
-\frac4{\pi^2}\int_{\gamma_1}\int_{\gamma_2}\frac1{(\overline{z_2}-\overline{z_1})^2}
d\overline{z_1}d\overline{z_2} .$$

The first of these integrals gives
$$-\frac4{\pi^2}\log\frac{(p-q)(r-s)}{(p-s)(r-q)}.$$
Therefore
\begin{eqnarray*}
\E((h(p)-\bar h(p))(h(q)-\bar h(q)))&=&\frac4{\pi^2}\left(
-2\Re\log\frac{(p-q)(r-s)}{(p-s)(r-q)}+2\Re\log\frac{(\overline{p}-q)(r-s)}{
(\overline{p}-s)(r-q)}\right)\\
&=&\frac8{\pi^2}\Re\log\left(\frac{\overline{p}-q}{p-q}\right).
\end{eqnarray*}

\section{Trees and winding number.}
\label{trees}
A {\bf directed spanning tree} on a (undirected) graph $G$ is a 
connected contractible (acyclic) collection of edges of $G$, where each edge has
a chosen direction such that each vertex but one
has exactly one outgoing edge. The single vertex with no outgoing
edge is called the {\bf root} of the tree.
If $G$ is a graph with boundary, (that is, there is a subset of 
vertices called the {\bf boundary} of $G$), then a {\bf directed essential
spanning forest} is a collection of edges of $G$, each component of which is
contractible, where each edge has a chosen direction,
such that each non-boundary vertex has
exactly one outgoing edge, and no boundary vertex has an outgoing
edge. 

\begin{figure}[htbp]
\begin{center}
\PSbox{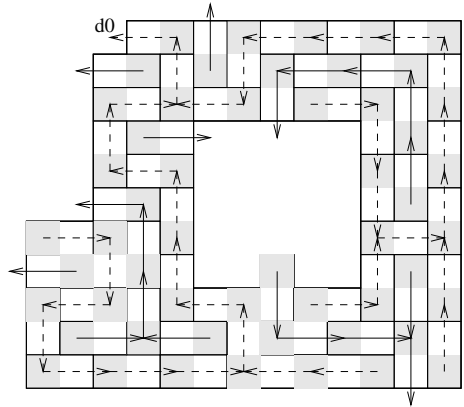}{1in}{2in}
\end{center}
\caption{\label{tree}The directed essential spanning forest (solid arrows)
associated to the tiling of Figure \protect\ref{hts}. The dual tree
is shown in dotted arrows.}
\end{figure}

``Temperley's trick" (see \cite{BP})
is a mapping between domino tilings
of certain polyominos and directed essential spanning forests 
of associated graphs.
In the case $P$ is a Temperleyan polyomino,
the directed essential spanning forest is on the graph $\Bo'(P)$
of section \ref{M'} and the boundary consists of the set
$Y$.  The forest is defined from a tiling as follows. 
Each square $v$ in $B_0\cap P$ is covered by a domino.
The white square of this domino lies over an edge of $\Bo'(P)$.
This edge is chosen to be the outgoing edge of $v$ on
the tree on $\Bo'(P)$.  See Figure \ref{tree} for the directed essential
spanning forest associated to the domino tiling of Figure \ref{hts}.

To see that the essential spanning forest constructed from a tiling
has no cycles, it suffices to construct
the planar dual forest, which is constructed in a similar way
from the graph $\Bl(P)\cup\{d_0\}$.
In the case $P$ is a Temperleyan polyomino,
the dual forest is a tree rooted at $d_0$ (since $d_0$ is the only possible root).
Since the dual tree is connected
the primal tree has no cycles.

Conversely, any essential spanning forest on $\Bo'(P)$ gives a domino tiling of $P$,
so these systems are in bijection.

The height function of a domino tiling has a nice interpretation 
for the directed paths in the associated spanning tree. 
To a vertex $v$ in $\Bo'(P)$ associate a height which is the average
of the heights of the four vertices of $P$ adjacent to $v$.
If the outgoing edge of the tree at $v$ points to an adjacent
vertex $v'$, and the outgoing edge at $v'$ points to
a vertex $v''$, then the height at $v'$ equals the
height at $v$ if the three vertices $v,v',v''$ are aligned;
if the path turns left at $v'$ then the height at $v'$
is one less than the height at $v$; if the path turns
right at $v'$ then the height at $v'$ is one more
than the height at $v$.

Therefore the height function along the directed path 
measures the net turning of the path.

\begin{prop} Let $P$ be a Temperleyan polyomino with a tiling
and let $T$ be the associated essential spanning forest.
The height change along a directed path $\gamma$ in $T$
equals the net turning of the
path, that is, the number of right turns minus the number
of left turns.
\end{prop}

In particular if $\gamma$ is a directed
path in $T$ running between $d_j\in D_j$ and the outer boundary,
the height difference between $D_j$ and $D_0$ 
is exactly measured by the winding number of the
path $\gamma$ (around $D_j$).

\begin{figure}[htbp]
\begin{center}
\PSbox{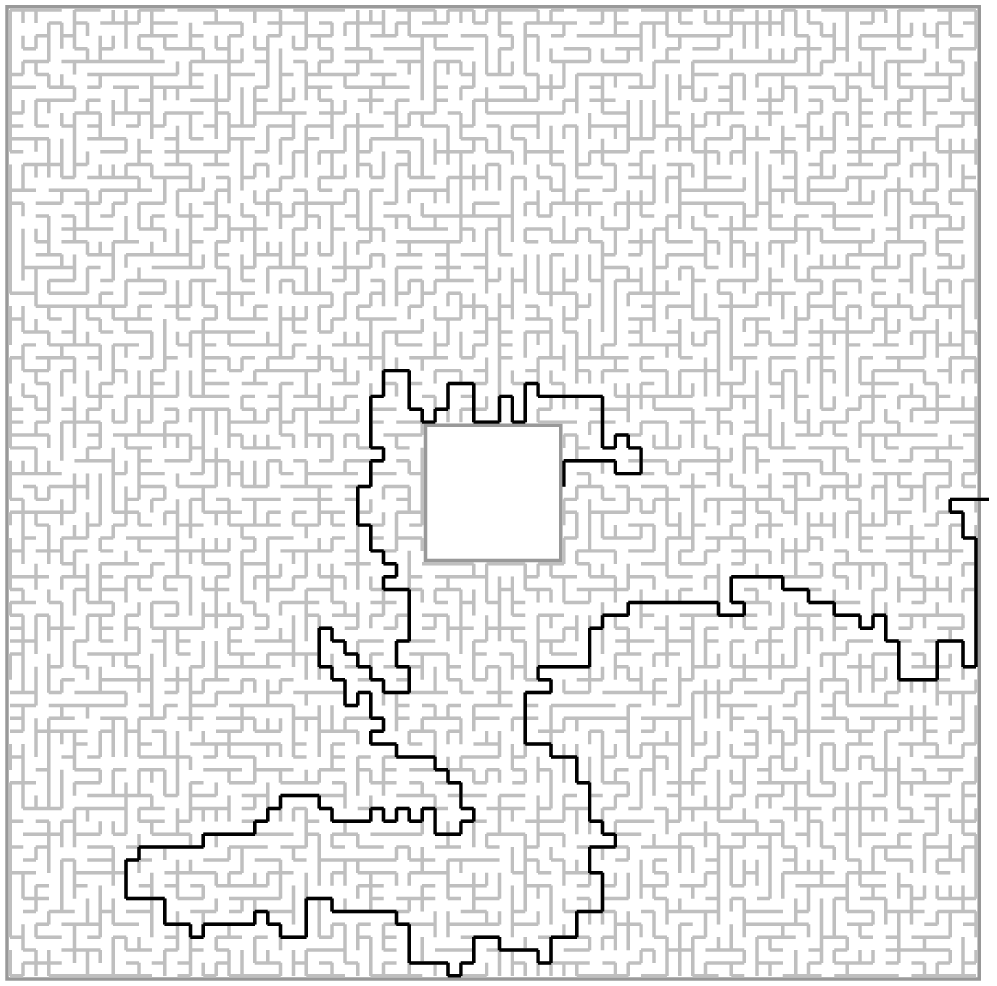}{1in}{3.5in}
\end{center}
\caption{\label{treefig}The spanning forest associated to 
a tiling of an annulus.}
\end{figure}

In Figure \ref{treefig}
we show the spanning tree associated to a tiling of
a Temperleyan annulus in which the height difference between the boundaries
is $4$. The directed path from a vertex adjacent to
$d_1$ to $d_0$ is highlighted.
\section{Other boundary conditions}\label{conclusion}
There are a number of intuitive ideas in the proof  of Theorem \ref{1}
which are worthwhile exploring. Foremost is the interesting
link between the height function along a boundary component and the
singularities of the coupling function. When we introduced 
the exposed vertices in our polyominos (in order to
make it tilable) we `created' poles in the coupling function
at those points.
There are a number of other, equally simple, boundary
conditions which give different boundary behavior
for the coupling function. The most natural seems to be to have
all boundary edges have even length. This is natural
from the point of view of tilings since it is
trivial to show that such a region has a tiling. Furthermore
the height function along such a boundary is particularly simple in this case. 
However the boundary conditions for the coupling function are more
difficult: on some boundary edges the real part will be zero and on
others the imaginary part will be zero. The coupling function
will have poles at certain corners and zeros at the remaining corners.
It seems more difficult to prove the convergence
of the coupling function when $\eps\to0$ in this case.

Another potential improvement in the proof would be 
a more general result (more general than Corollary \ref{bdygreen}) 
concerning the convergence of
the discrete Green's function centered near the boundary of a domain.
Surprisingly, this problem does not seem to have been considered
in the literature.

Another direction to be explored is the case of regions without boundary.
In \cite{Kenyon} we computed a formula for the coupling function
on a torus. By a recent result of Tesler \cite{Tesler} higher-genus
surfaces can be handled by similar methods.

\bigskip

\noindent
CNRS UMR8628, Laboratoire de Topologie, B\^at. 425,
Universit\'e Paris-Sud, 91405 Orsay, France.
\end{document}